\documentclass[manuscript]{acmart}

\usepackage{listings}

\usepackage{algorithm}
\usepackage{algorithmic}

\usepackage{subfig}

\usepackage{tabu}
\usepackage{multirow}

\AtBeginDocument{%
	\providecommand\BibTeX{{%
			\normalfont B\kern-0.5em{\scshape i\kern-0.25em b}\kern-0.8em\TeX}}}

\begin{document}
\title{FUBOCO: Structure Synthesis of Basic Op-Amps by FUnctional BlOck COmposition}

\author{Inga Abel}
\email{inga.abel@tum.de}
\orcid{0000-0002-9996-4162}
\author{Helmut Graeb}
\email{helmut.graeb@tum.de} 

\affiliation{%
        	\institution{Technical University of Munich, Chair of Electronic Design Automation}
        	\streetaddress{Arcisstr. 21}
        	\city{Munich}
        	\country{Germany}}

\begin{abstract}
This paper presents a method to automatically synthesize the structure of an operational amplifier. It is positioned between approaches with fixed design plans and a small search space of structures and approaches with generic structural production rules and a large search space with technically impractical structures. The presented approach develops a hierarchical  composition graph based on functional blocks that spans a search space of thousands of technically meaningful structure variants for single-output, fully-differential and complementary operational amplifiers. The search algorithm is a combined heuristic and enumerative process. The evaluation is based on circuit sizing with a library of behavioral equations of functional blocks. Formalizing the knowledge of functional blocks in op-amps for structural synthesis and sizing inherently reduces the search space and lessens the number of created topologies  not fulfilling the specifications. Experimental results for the three op-amp classes are presented.  An outlook how this method can be extended to multi-stage op-amps is given.
\end{abstract}


\maketitle

\section{Introduction} 
Automation of structural synthesis of analog circuits has not gained much attention from industry. A reason for this might be that the current approaches are following two major paths, none of which seems to be attractive from the industrial point of view. 
One path of structural synthesis is a specification-based selection of one or few netlists from a library \cite{IntegerBasedTopologieSelecting, NovelCircuitTopologySynthesisMethodUsingCircuitFeatureMiningAndSymbolicComparison, AutomaticTopologySelectionAndSizingOfClassDLoopFiltersForMinimizingDistortion, AGenericTopologySelectionMethodForAnalogCircuits,HARC89,ALowVoltageLowPowerAmplifierCreatedByReasoningBasedSystematicTopologySynthesis}. This is an automation of what happens in practice, as every company has a set of, say, 20 to 30 different netlists for op-amps, from which is initially chosen according to the specification. As an experienced designer can do this selection instantly, there is little gain in design time or design quality. The other path of structural synthesis builds up the netlist by combining modules of transistors and transistor groups while satisfying Kirchhoff laws and basic voltage/current conversion at the module interfaces \cite{AnAutomatedTopologySynthesisFrameworkForAnalogIntegratedCircuits, AGraphGrammarBasedApproachToAutomatedMultiObjectiveAnalogCircuitDesign, FEATS, VariationAwareStructuralSynthesisOfAnalogCircuitsViaHierarchicalBuildingBlocksAndStructuralHomotopy, StructuralSynthesisAndOptimizationOfAnalogCircuits,AutomaticSynthesisAndSizingGeneticProgramming, Encoding}. Here, a plethora of variants is created, from which the promising ones are selected only after symbolic analysis and complete sizing. Designers see an unnecessary variety of intermediate solutions that they would never have  created. In the presented approach, a new path for analog structural synthesis is started, which lies in the middle between these two poles. It does more than selecting from 20 to 30 available alternatives. And it creates a much smaller number of intermediate solutions that fail the requirements.

Early topology synthesis approaches were \cite{IntegerBasedTopologieSelecting, HARC89, StructuralSynthesisAndOptimizationOfAnalogCircuits}. While \cite{StructuralSynthesisAndOptimizationOfAnalogCircuits} was tested on different capacitor structures, as LC-structures or switch-capacitors, \cite{IntegerBasedTopologieSelecting, HARC89} were developed for operational amplifiers. In \cite{HARC89}, a top level abstraction of a circuit is implemented and hierarchically transformed into a transistor level structure with a fixed design plan for evaluation. \cite{IntegerBasedTopologieSelecting} has a  library of 64 topologies and fixed design plans.
Synthesis approaches based on genetic algorithms followed to overcome the topology dependence of the early approaches \cite{VariationAwareStructuralSynthesisOfAnalogCircuitsViaHierarchicalBuildingBlocksAndStructuralHomotopy, AutomaticTopologySelectionAndSizingOfClassDLoopFiltersForMinimizingDistortion, AutomaticSynthesisAndSizingGeneticProgramming, Encoding}. They are based on libraries containing simple transistor structures. With these methods, many different circuit topologies are created.  However, many topologies are redundant.
To size the topologies and evaluate  their practicability, simulations are used.
To lessen the number of redundant op-amp topologies, graph-grammar based approaches were developed \cite{AnAutomatedTopologySynthesisFrameworkForAnalogIntegratedCircuits, AGraphGrammarBasedApproachToAutomatedMultiObjectiveAnalogCircuitDesign, FEATS}. With strict grammar rules and isomorphism techniques, the number of redundant topologies are reduced. The evaluation and sizing of the topologies takes place after synthesizing all possible topologies. Thus, much computation time is spent for circuits which cannot fulfill the given specifications from a visual inspection of a designer.
Other approaches to lessen the number of created topologies were rule-based topology synthesis algorithms which closely implemented designer knowledge \cite{NovelCircuitTopologySynthesisMethodUsingCircuitFeatureMiningAndSymbolicComparison,AGenericTopologySelectionMethodForAnalogCircuits, ALowVoltageLowPowerAmplifierCreatedByReasoningBasedSystematicTopologySynthesis}.

Different to previous methods, the presented synthesis approach contains a comprehensive computer-oriented systematic of op-amp functional building blocks like, e.g., amplification stages with their internal transconductances, loads and biases. The functional blocks  form a composition graph.
Thus, it overcomes the  topology dependency in \cite{IntegerBasedTopologieSelecting, HARC89, AGenericTopologySelectionMethodForAnalogCircuits}. But it is still close enough to the op-amp structure such that  redundant or impracticable topologies are avoided. 
This makes algorithms to remove these structures as in \cite{AnAutomatedTopologySynthesisFrameworkForAnalogIntegratedCircuits, FEATS} unnecessary.
A novel algorithm for structural synthesis iteratively composes op-amp netlists from these basic functional blocks. For given specifications, the netlists are evaluated using an analytical equation-based sizing method similar to \cite{ABNG20b}. A library was developed storing analytical behavior models for every functional block in an op-amp \cite{ABNG20c}. Useless configurations are avoided in an early phase of the structural synthesis process. For instance, there exist performance features which degrade if a second stage is added to a one-stage op-amp topology, e.g., area, phase margin. In this case, the corresponding two-stage op-amp topology is not created if the one-stage op-amp already indicates to fail these specifications. Introducing design equations into the structural synthesis process obviates the need for symbolic analysis \cite{StructuralSynthesisAndOptimizationOfAnalogCircuits} or circuit simulation \cite{VariationAwareStructuralSynthesisOfAnalogCircuitsViaHierarchicalBuildingBlocksAndStructuralHomotopy,Reviewer1} in evaluating intermediate solutions. This is achieved by a rigorous adoption of the practical creation process of analog designers along the functional block hierarchy of an operational amplifier. 

This paper is an extended version of \cite{ICCAD}. \cite{ICCAD} presented a prototype of the synthesis tool, creating up to 34 op-amp topologies on amplification stages and evaluating the topologies through sizing. In this paper,  we present the functional block composition of several thousand op-amp variants, building up the amplification stages by their subblocks, starting on transistor level with basic one-transistor functional blocks.
New compared to \cite{ICCAD}, it presents:
\begin{itemize}
	\item An algorithm to synthesize every functional block of an op-amp by its functional subblocks starting on transistor level up to whole op-amp topology (Sec. \ref{sec:functionalBlockSynthesis}).
	\item An algorithm to synthesize a functional block based bias circuit for an op-amp topology (Sec. \ref{sec:BiasSynthesis}).
	\item An structural synthesis algorithm for op-amps featuring thousands of variations (Sec. \ref{sec:synthesisSimpleOpAmp} and Sec. \ref{sec:synthesisAlgorithm}).
	\item Experimental results (Sec. \ref{sec:experimentalResults}) featuring three op-amp types: single-output, fully-differential, complementary.
	\item An outlook how this method can be extended to three-stage op-amp topologies (Sec. \ref{sec:OutlookThreeStageOpAmp}).
\end{itemize}
With the enhanced version of the synthesis method, up to 3912 op-amps can now be synthesized. The number of created op-amps  depends on the type of specification. Broad specifications result in many synthesized topology alternatives, while strict specifications lead to a small set of synthesized netlists.

The goal of this approach is to perform op-amp synthesis using a formalized, computer-oriented description of the fundamentals of op-amp design. The hierarchical structure of the approach allows a straightforward extension to further functional blocks and multi-stage op-amp architectures. This allows not only to create suitable op-amps for given specifications sets, but also to create large circuit libraries containing only practicable solutions. Hence, the created libraries provide large data sets for future machine learning projects.

In the following, an overview of the functional blocks in op-amps is given (Sec. \ref{sec:functionalBlocks}). Sec. \ref{sec:functionalBlockSynthesis} describes the generic algorithm to synthesis each functional block in Sec. \ref{sec:functionalBlocks} based on its subblocks. Sec. \ref{sec:BiasSynthesis} describes the synthesis of the op-amp bias circuit. Sec. \ref{sec:synthesisSimpleOpAmp} describes the creation of whole op-amp topologies based on the previous described algorithms. The whole FUBOCO synthesis process is presented in Sec. \ref{sec:synthesisAlgorithm}. Experimental results featuring three op-amp types are presented in Sec. \ref{sec:experimentalResults}. Sec. \ref{sec:OutlookThreeStageOpAmp} discusses the extension of the method to three-stage op-amps. The paper ends with a conclusion (Sec. \ref{sec:Conclusion}).

\section{Functional Blocks in Op-Amps}\label{sec:functionalBlocks}
\begin{figure*} [tb] \centering 
		\subfloat[Folded-cascode op-amp with common mode feedback (CMFB) (Fully-differential, one-stage)]{
			\label{fig:foldedCascodeOpAmp}
			\includegraphics[width=0.51\linewidth]{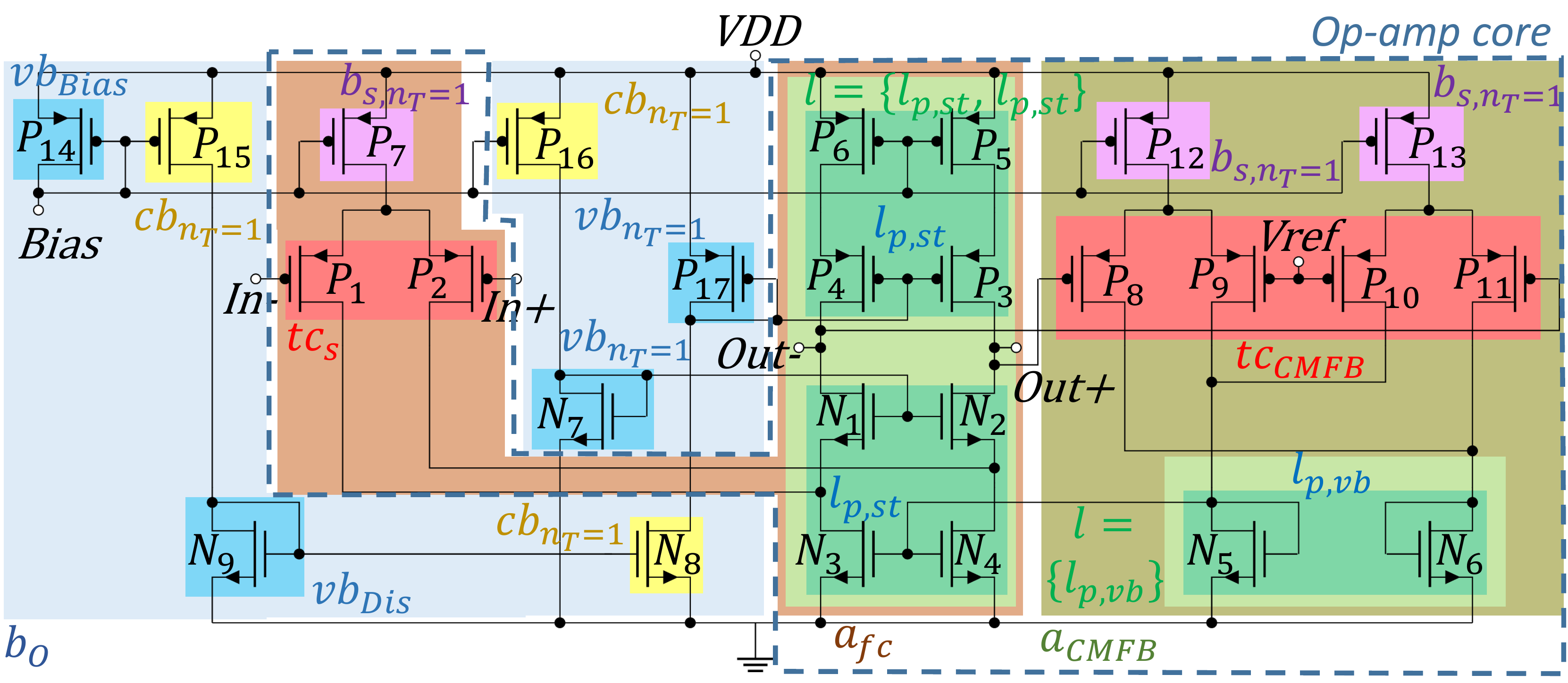}
		}% 
		\qquad
		\subfloat[Symmetrical op-amp with cascode second stage (Single-output, symmetrical)]{
			\label{fig:symmetricalOpAmp}
			\includegraphics[width=0.32\linewidth]{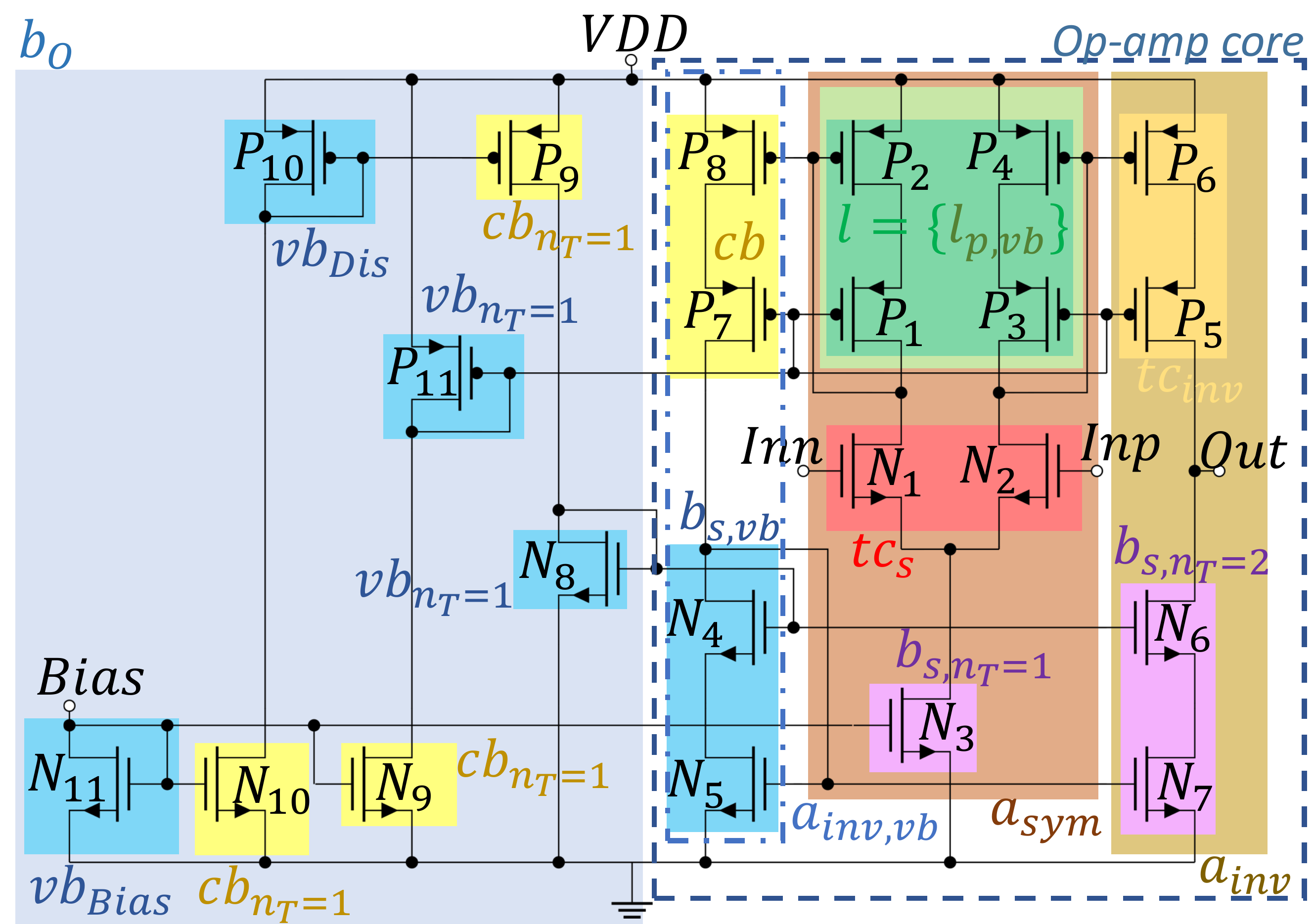}
		}% 
		\qquad
		\subfloat[Complementary op-amp (One-stage)]{
			\label{fig:railToRailAmplifier}
			\includegraphics[width=0.47\linewidth]{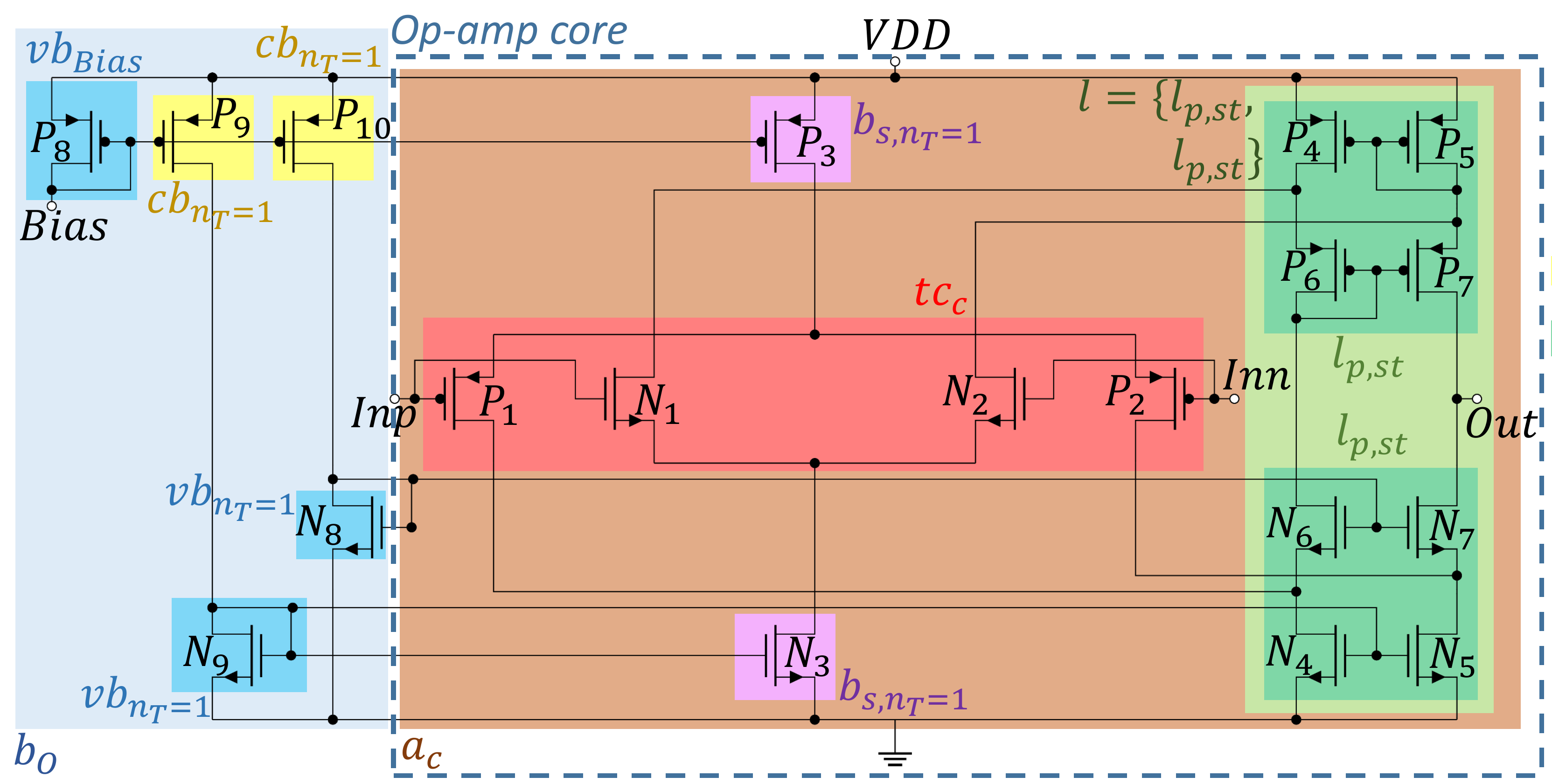}
		} %
		\qquad	
		\subfloat[Three-stage op-amp (Outlook, Sec. \protect \ref{sec:OutlookThreeStageOpAmp}) ]{
			\label{fig:threeStageOpAmp}
			\includegraphics[width= 0.4\linewidth]{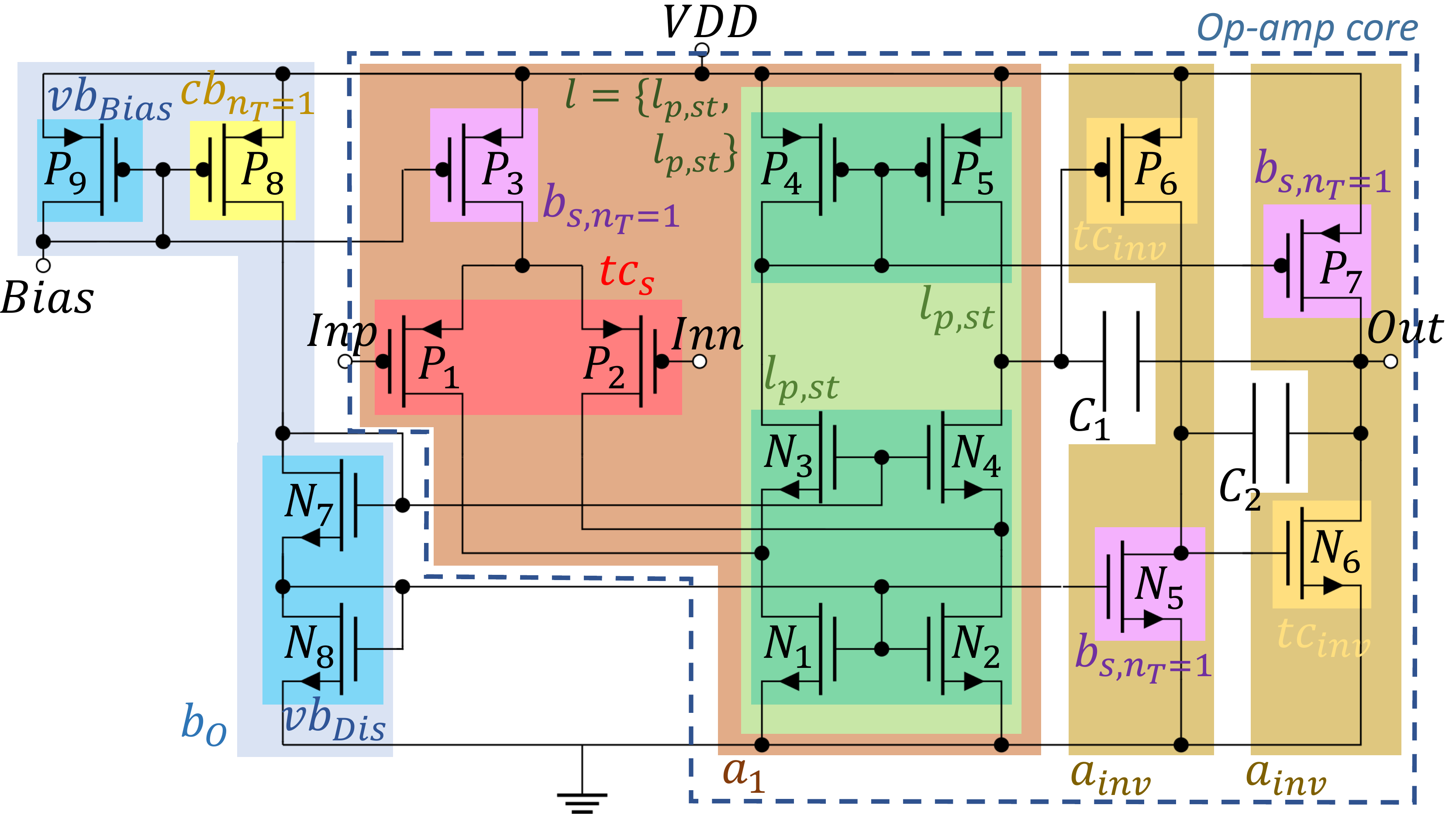}
		}%	
		\qquad
		\caption{Example topologies synthesizable with the presented method, colored background: functional blocks of HL 2 - 4}
		\label{fig:differentOpAmpToplogies}
\end{figure*}

Every op-amp consists of a set of transistor blocks which can be characterized by their function and are therefore called functional blocks in the following. This paper presents a hierarchical structuring of these functional blocks and a computer-oriented representation of the fundamental principles of op-amp design composition as described in standard works~\cite{Allen,LakerSansen, AnalogIntegratedCircuitDesign, GrayMeyer, Palumbo, MOSCapacitances}.
An overview of the basic functional blocks structured on five hierarchical levels is given in~Fig.~\ref{fig:FunctionalBlockLibrary}. 
Examples are shown using four different op-amp topologies in Fig.~\ref{fig:differentOpAmpToplogies}.
A more detailed description is given in \cite{ABNG20}.

\begin{figure}
\centering 
\setlength{\tabcolsep}{0.1cm}
\begin{tabular}{lll}	
		HL 5 & Op-amps & single-output (SO), fully-differential (FD), complementary (Comp) \\\hline	
		HL 4 & Op-amp subblocks & op-amp bias ($b_O$), amplification stage ($a$) \\\hline
		HL 3 & Amplification stage subblock & load ($l$), transconductance ($tc$), stage bias ($b_s$) \\\hline
		HL 2 & Structures & voltage bias ($vb$), current bias ($cb$), differential pair ($dp$),  \\\hline
		HL 1 & Devices & normal transistor ($nt$), diode transistor ($dt$), capacitor ($cap$)\\ 	
	\end{tabular}
	\caption{Hierarchical structure of functional blocks in op-amps}\label{fig:FunctionalBlockLibrary}	
\end{figure}

\emph{Hierarchy level 1} (HL 1) consists of devices, e.g., {\em capacitors cap} and transistors. Transistors are further divided into groups according to their self-connection.
{\em Normal transistors} ($nt$) are transistors without any self-connection, e.g., Fig. \ref{fig:foldedCascodeOpAmp}, $P_1$. {\em Diode transistors} ($dt$) are transistors with their gates connected to their drains, e.g., Fig. \ref{fig:foldedCascodeOpAmp} $P_{14}$.

{\em Hierarchy level 2} (HL 2) consists of transistor structures, e.g., {\em voltage bias} ($vb$, Fig. \ref{fig:foldedCascodeOpAmp} $vb_{Bias},vb_{Dis}$), {\em current bias} ($cb$, Fig. \ref{fig:foldedCascodeOpAmp} $cb_{n_T = 1}$) and {\em differential pair} ($dp$, Fig. \ref{fig:symmetricalOpAmp} $N_1,N_2$).
The voltage and current biases can be simple, e.g. Fig. \ref{fig:foldedCascodeOpAmp} $vb_{n_T=1}$, or cascode (Fig. \ref{fig:threeStageOpAmp} $vb_{Dis}$).

\emph{Hierarchy level 3 } (HL 3) consists of the amplification stage subblocks which are the {\em transconductance} ($tc$, Fig.~\ref{fig:differentOpAmpToplogies}, red), the  {\em in-stage load} ($l$, Fig. \ref{fig:differentOpAmpToplogies}, light green) and the {\em stage bias} ($b_s$, Fig.~\ref{fig:differentOpAmpToplogies}, violet). In the following, we abbreviate the in-stage load to load.
For the transconductance, two main  types exist: non-inverting ($tc_{ninv}$, Fig.~\ref{fig:symmetricalOpAmp} $tc_s$) and inverting ($tc_{inv}$, Fig.~\ref{fig:symmetricalOpAmp}).
The non-inverting transconductance can be further divided into three types: simple ($tc_s$, Fig. \ref{fig:symmetricalOpAmp}), complementary ($tc_c$, Fig.~\ref{fig:railToRailAmplifier}),  and common-mode feedback (CMFB) ($tc_{CMFB}$, Fig. \ref{fig:foldedCascodeOpAmp}).  The load consists of one or two load parts ($l_p$, Fig.~\ref{fig:differentOpAmpToplogies}, dark green).

\emph{Hierarchy level 4} (HL 4) consists of op-amp subblocks, which are the {\em amplification stages} ($a$), and the {\em op-amp bias} ($b_O$).
Two types of amplification stages  exist: non-inverting ($a_{ninv}$, Fig.~\ref{fig:foldedCascodeOpAmp} $a_{fc}$), and inverting ($a_{inv}$, Fig. \ref{fig:symmetricalOpAmp}).  The non-inverting amplification stages are further divided into simple first stage ($a_s$), complementary first stage ($a_c$, Fig.~\ref{fig:railToRailAmplifier}), telescopic first stage ($a_{tel}$), folded-cascode first stage ($a_{fc}$, Fig.~\ref{fig:foldedCascodeOpAmp}), symmetrical first stage  ($a_{sym}$, Fig. \ref{fig:symmetricalOpAmp}) and common-mode feedback stage ($a_{CMFB}$, Fig. \ref{fig:foldedCascodeOpAmp}).

\emph{Hierarchy level 5} (HL 5) consists of the op-amp itself.
Three types of op-amps are subject of this paper: single-output (SO), fully-differential (FD)  and complementary (Comp). Complementary op-amps are defined as op-amps having a pmos and nmos differential pair forming the input stage. 

In the following, we will distinguish between the $core$ of an op-amp, containing the amplification stages and the capacitors,  and the {\em op-amp bias} $b_O$, containing all transistors needed to bias the structures in the op-amp core (Fig. \ref{fig:differentOpAmpToplogies}).

\section{Synthesis of Functional Blocks Except Op-Amp Bias} \label{sec:functionalBlockSynthesis}
The hierarchical structure of functional blocks allows the synthesis of structural implementations of a functional bock based on its subblocks.
This section presents a new generic algorithm  which covers all blocks in Fig. \ref{fig:FunctionalBlockLibrary} except the op-amp bias $b_O$. A separate algorithm to synthesize the op-amp bias is presented in Sec. \ref{sec:BiasSynthesis}.

\begin{figure} [tp] \centering
	\includegraphics[width=  0.7\linewidth]{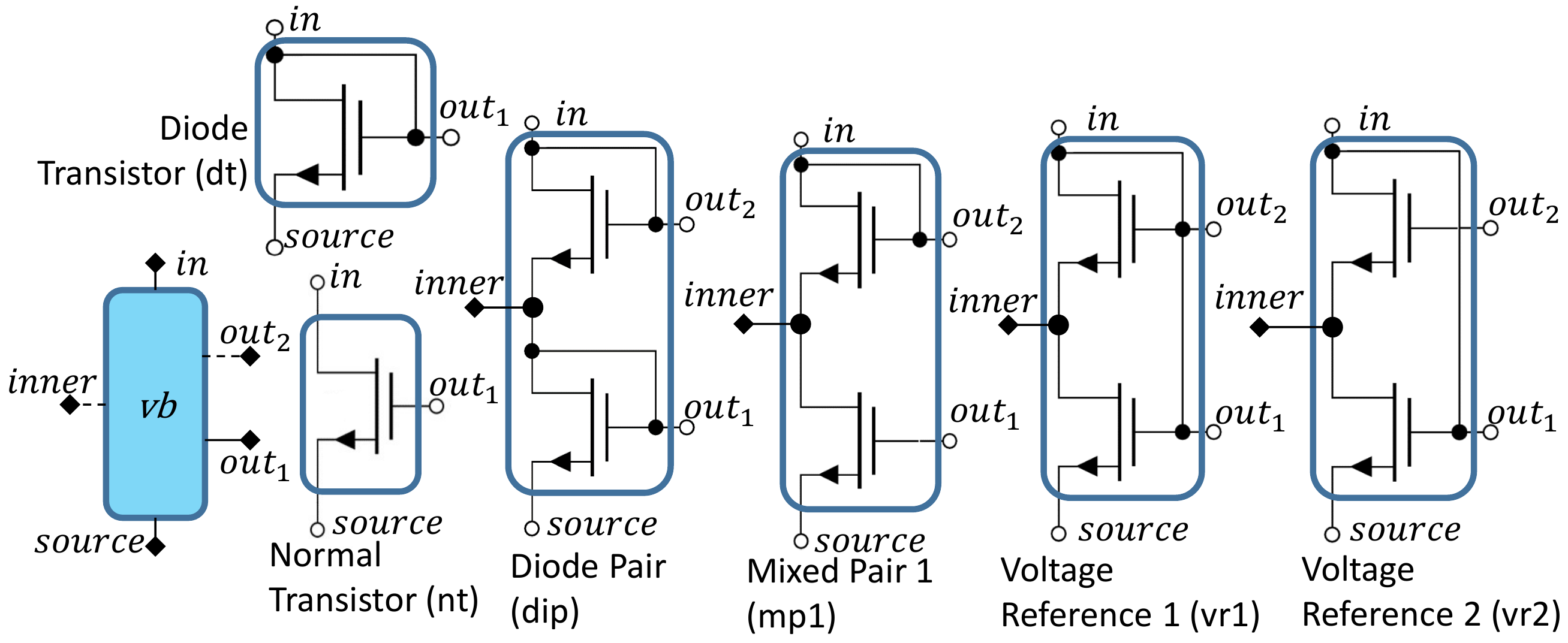}
	\caption{Voltage bias instance and corresponding structural implementations}
	\label{fig:voltageBiasStrucImp}
\end{figure}

\subsection{{Data Structure}}\label{sec:dataStructure}
A generic approach to synthesis of op-amp functional blocks requires the transition from a functional, i.e., behavioral, description of a block, to a transistor implementation, i.e., to a structural description, of a block. In this transition, different and new pins may arise. In this work, this transistion is implemented by representing functional blocks as instances with specific sets of pins. The set of pins varies for different functional block (implementation) types.  Each functional block type has its own specific set of pins.
The functional block {\em voltage bias} for instance has two implementation types, i.e., simple and cascode (Fig.~\ref{fig:voltageBiasStrucImp}).
If the voltage bias is simple, its instance has three pins: $in, out_1, source$. If the voltage bias is cascode, its instance has two additional pins: $inner, out_2$. The pins define the generic pin sets for the two implementation types. They cover all possible implementation sets, even if an implementation as for instance {\em vr1} happens to connect two pins.  This data structure provides  exchangeability and flexibility in the synthesis process.

\begin{figure}[]\centering
	\setlength{\tabcolsep}{0.05cm}
	{
		\begin{tabu}{ll}
			${S}_1, \ldots, {S}_i$:& Structural implementation sets of FB$_1, \ldots$, FB$_i$  ($2, \ldots, i$ optional)\\
			${R}_c$ (optional):& Characteristic connections of FB$_{new}$\\
			${\mathcal{R}}_f$ (optional):& Rules $s_{new}$  must fulfill\\
			${R}_a$ (optional):& Additional connections $s_{new}$ can have\\
			${P}_{\text{FB}_{new}}$:& Pin set of $s_{new}$ 
	\end{tabu}}
	\caption{Input of Algorithm \protect \ref{algo:SynthesisFunctionalBlock}, FB$_j$: $j$th functional block, $s_j$: a structural implementation of FB$_j$ }\label{tab:InputAlgorithm}
\end{figure}

\subsection{Generic Algorithm to Synthesize a Functional Block (Except Op-Amp Bias)}
Alg. \ref{algo:SynthesisFunctionalBlock} creates for every functional block FB$_{new}$ in Fig.~\ref{fig:FunctionalBlockLibrary} (but the op-amp bias) a set of  structural implementations (instances) ${S}_{new}$. 

\subsubsection{Input}

The input of the algorithm is defined in Fig.~\ref{tab:InputAlgorithm} and comprises the following sets. 

{\em Structural implementations} ${S}_{1}, {S}_2, .., {S}_i$ of the functional subblocks of FB$_{new}$: 
$i$ is the number of functional subblocks that are combined to build FB$_{new}$. E.g., a cascode voltage bias in
Fig.~\ref{fig:voltageBiasStrucImp} consists of two functional blocks, hence there are two
sets $S_1$, $S_2$. Each set consists of normal  and diode  transistors $nt \in NT, dt \in DT$ having the same doping $\Phi$ (${S}_1:NT_\Phi, DT_{\Phi}; {S}_2: NT_\Phi, DT_{\Phi};$). Further input defines which combinations of these are allowed in a cascode voltage bias.

{\em Characteristic connections} ${R}_c$ state how the instances in ${S}_1,.., {S}_i$ are connected to each other. This input is optional as no characteristic connections are provided for functional blocks consisting of one instance, e.g., simple voltage bias. To create a cascode voltage bias, $R_c$ contains that the drain of a transistor $s_1 \in S_1$ must be connected to the source of a transistor $s_2 \in S_2$ (${R}_c: s_1.drain \leftrightarrow s_2.source$). %Thus every combination of a transistor $s_1 \in S_1, s_2 \in S_2$ forms a transistor stack.

{\em Functional block rules} ${\mathcal{R}}_f$ define wanted and unwanted connections independent of the characteristic connections. This is used to verify that a combination of different instances of functional subblocks is an implementation of FB$_{new}$.  E.g., ${\mathcal{R}}_f$ of a cascode voltage bias (Fig. \ref{fig:voltageBiasStrucImp}) contains that the pin $in$, i.e., $s_2.drain$, must be connected to one of the gates of its subblocks (${\mathcal{R}}_f: s_2.drain \leftrightarrow (s_1.gate \vee s_2.gate)$). 

${\mathcal{R}}_f$ also contains {\em basic structural rules} of analog building blocks as, e.g., that no transistor drain $t_{m}.drain$ is allowed to be connected to another transistor drain $t_{n}.drain$ of the same doping $\Phi$:
\begin{equation}\label{eq:basicRule}
	\forall_{t_m, t_n \in T_{\Phi}} t_{m}.drain \nleftrightarrow t_{n}.drain
\end{equation}
$T_{\Phi}$ are all transistors in the newly created implementation $s_{new}$ of FB$_{new}$ with doping type $\Phi$.
If the diode pair ($dip$) in Fig.~\ref{fig:voltageBiasStrucImp} would have a connection between $out_1$ and $out_2$ it would not be a valid structural implementation of a voltage bias.

{\em Additional connections} ${R}_a$ formulate additional optional connections to the connections in ${R}_c$. 
In the $vr1$-implementation of a cascode voltage bias (Fig. \ref{fig:voltageBiasStrucImp}), the gates of the transistor are additionally connected. In the voltage reference 2, the gate of the lower transistor is additionally connected to the drain of the transistor above (${R}_a: \{s_1.gate \leftrightarrow s_2.gate\}, \{s_2.drain \leftrightarrow s_1.gate\}$).

\begin{algorithm} [tp]{
		\caption{{ Synthesis of a functional block except op-amp bias}} \label{algo:SynthesisFunctionalBlock}
		\begin{algorithmic}[1]
			\REQUIRE Compare Fig. \ref{tab:InputAlgorithm} 
			\STATE ${S}_{new}$ := $\{ ~~\}$ //{ The set of structural implementations of {FB}$_{new}$ is empty}
			\FORALL{$s_1 \in {S}_1$}
			\FORALL{$s_2 \in {S}_2$}
			\STATE ....
			\FORALL {$s_i \in {S}_i$}
			\STATE $c_{new}$ := createConnections($s_1, s_2, ..., s_i, {R}_c$)
			\IF{fulfillesRules(${\mathcal{R}}_f$,$c_{new}$)} 
			\STATE $s_{new}$ := createNewInstance($c_{new}, {P}_{fb_{new}}$)
			\STATE ${S}_{new}$ := ${S}_{new} \cup \{s_{new}\}$
			\ENDIF
			\FORALL{$r_a \in {R}_a$}
			\STATE $c_{new}$ := createConnections($c_{new}, r_a$)
			\IF{fulfillesRules(${\mathcal{R}}_f$,$c_{new}$)} 
			\STATE $s_{new}$ := createNewInstance($c_{new}, {P}_{fb_{new}}$)
			\STATE ${S}_{new}$ := ${S}_{new} \cup \{s_{new}\}$
			\ENDIF
			\ENDFOR
			\STATE ...
			\ENDFOR
			\ENDFOR
			\ENDFOR
			\RETURN ${S}_{new}$
	\end{algorithmic}}
\end{algorithm}

\subsubsection{Algorithm}
To synthesize all structural implementations of a functional block, 
the algorithm iterates over the sets of  structural implementations ${S}_1, {S}_2, .., {S}_i$ to create all possible combinations.
For each combination, a subcircuit $c_{new}$ is created, consisting of  $s_1 \in {S}_1$, $s_2 \in {S}_2$, ...,  $s_i \in{S}_i$ having the required connections in ${R_c}$.
It is checked if $c_{new}$ fulfills the rules in ${\mathcal{R}}_f$. If that is the case, a new instance $s_{new}$ is created being a valid structural implementation of FB$_{new}$.
To create all implementations of a cascode voltage bias, every combination of diode and normal transistor with same doping having a drain-source connection is created. Because of $R_{f}$, only the diode pair and the mixed pair are recognized as valid structural implementations.

If a set ${R}_a$ is provided, subcircuits having the defined additional connections are also created. If these circuits fulfill 
${\mathcal{R}}_f$, the respective instances are created. Voltage reference 1 and voltage reference 2 are thus created as cascode voltage bias. Further transistor structures are created based on $R_{a}$, but are discarded as they do not fulfill $\mathcal{R}_{f}$.

\section{Synthesis of the Op-Amp Bias}\label{sec:BiasSynthesis}
\begin{figure} [tp] \centering
	\includegraphics[width=  0.6\linewidth]{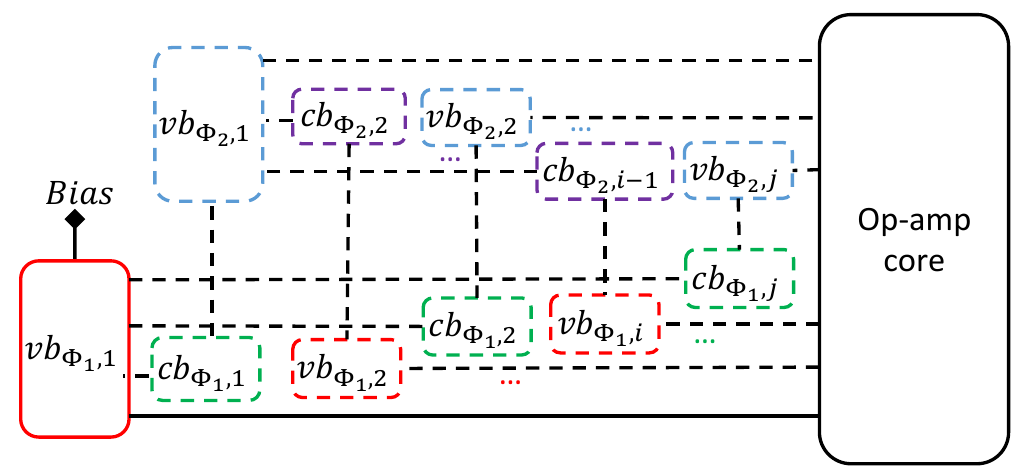}
	\caption{Schematic overview of a bias}
	\label{fig:SchematicBias}
\end{figure}

While Alg. \ref{algo:SynthesisFunctionalBlock} is used to synthesize the {\em op-amp core}, i.e., the amplification stages and capacitor structures with their connections, the op-amp bias $b_O$ is created with Alg. \ref{algo:SynthesisBias}.

\subsection{Structure of an Op-Amp Bias}

An op-amp bias $b_O$ consists of $n$ voltage biases ($vb$) and $n-1$ current biases ($cb$) \cite{ABNG20}. 
Its generic structure is defined in Fig.~\ref{fig:SchematicBias}.
The voltage biases supply the op-amp core with the required voltage potentials. The current biases supply these voltage biases with a current.  One voltage bias of each doping ($vb_{\Phi_1,1}$, $vb_{\Phi_2,1}$) supplies these current biases, called  {\em distributor voltage bias} ($vb_{Dis}$) in the following.
A single current input pin $p_{Bias}$  remains for the user-specified bias input current.
The number of voltage biases forming the bias depends on the position of the transistors in the op-amp core needing voltage supply. Five different position types are distinguished:
\begin{enumerate}
	\item {\em Improved Wilson current biases} are cascode current biases which have a diode transistor at the source and a normal transistor at the output. This type of current bias can only be connected to one implementation of a voltage bias (Fig.~\ref{fig:voltageBiasStrucImp} $mp1$). Together, they form an improved Wilson current mirror (Fig. \ref{fig:railToRailAmplifier} $P_4 - P_7$). For each  Wilson current bias in an op-amp, the specific voltage bias must be created (Alg.~\ref{algo:SynthesisBias}, Line \ref{line:ImprovedWilson}).
	\item {\em Cascode current biases} can be biased by a cascode voltage bias (Fig. \ref{fig:threeStageOpAmp}, $N_1 - N_4, N_7, N_8$).
	\item {\em Simple current biases with the source connected to the supply-voltage rail} are biased by simple voltage biases (Fig. \ref{fig:foldedCascodeOpAmp}, $P_7-P_{14}$)
	\item {\em Simple current biases  not connected to a supply-voltage rail} must be biased by an additional simple voltage bias. This is, e.g., the case in wide-swing cascode current mirrors (Fig.~\ref{fig:symmetricalOpAmp}, $N_4 - N_7$)). 
\end{enumerate}
A cascode current bias with doping $\Phi$ is not always biased by a cascode voltage bias. If additional single transistors of doping $\Phi$ are in the circuit needing voltage supply, the cascode current bias might be supplied by  two simple voltage biases (Fig. \ref{fig:railToRailAmplifier}, $N_3 - N_9$).

\begin{algorithm} [tbp]
	
	\caption{ Synthesis of op-amp bias} \label{algo:SynthesisBias}
	\begin{algorithmic}[1]
		\REQUIRE Set of transistors without voltage supply sorted according their doping and position in the op-amp core $T_{un}$ := $T_{un,1, \Phi_1} \cup T_{un,2, \Phi_1} \cup $ $T_{un,1, \Phi_2} \cup$ $T_{un,2, \Phi_2}$, Set of functional blocks forming the op-amp-core $op = \{a_1, ...\}$
		\STATE $VB_{\Phi_1}$ := $\{ ~~\}$ //The set of voltage biases of doping $\Phi_1$ 
		\STATE $VB_{\Phi_2}$ := $\{ ~~\}$ //The set of voltage biases of doping $\Phi_2$ 
		\STATE $VB_{\Phi_1,iw}, VB_{\Phi_2, iw}$:=createImprovedWilsonVoltageBiases($T_{un}$)\label{line:ImprovedWilson}
		\STATE $VB_{\Phi_1,add}$ := createAdditionalVoltageBiases($T_{un,1, \Phi_1}$, $T_{un,2, \Phi_1}$) //Alg.~\ref{algo:createVoltageBias}
		\STATE $VB_{\Phi_1}$ := $VB_{\Phi_1} \cup VB_{\Phi_1,iw} \cup VB_{\Phi_1,add}$
		\STATE $VB_{\Phi_2,add}$ := createAdditionalVoltageBiases($T_{un,1, \Phi_2}$, $T_{un,2, \Phi_2}$)
		//Alg. \ref{algo:createVoltageBias}
		\STATE $VB_{\Phi_2}$ := $VB_{\Phi_2} \cup VB_{\Phi_2, iw} \cup VB_{\Phi_2,add}$
		\STATE setBiasPin($VB_{\Phi_1}$, $VB_{\Phi_1}$) \label{line:BiasPin}
		\STATE $CB,VB_{dis}$ := createCurrentBiases($VB_{\Phi_1}, VB_{\Phi_2}$)//{Alg. \ref{algo:createCurrentBias}}
		\STATE $VB$ := $VB_{\Phi_1} \cup VB_{\Phi_2} \cup  VB_{dis} $
		\RETURN  $CB, VB$ 
	\end{algorithmic}
\end{algorithm}

\subsection{Generic Algorithm to Synthesize the Op-Amp Bias $b_O$}
Input of Alg. \ref{algo:SynthesisBias} are the transistors of the op-amp core needing voltage supply $T_{un}$ and the functional blocks of the op-amp core $op = \{a_1, ...\}$. The transistors in $T_{un}$ are sorted according to their doping $\Phi_1, \Phi_2$ and their position in the op-amp core. $T_{un,1}$ contains transistors connected with their source to a supply voltage rail, $T_{un,2}$ the remaining transistors.
To create the bias $b_O$ of the fully-differential op-amp (Fig.~\ref{fig:foldedCascodeOpAmp}), $T_{un, p, 1} = \{P_5, P_6, P_7, P_{12}, P_{13}\}$, $T_{un, p, 2} = \{P_3, P_4\}$, $T_{un, n, 1} = \{~~\}$, $T_{un, n, 2} = \{N_1, N_2\}$.

The algorithm creates and connects  the improved Wilson voltage  biases for all Wilson current biases in the op-amp core (Alg.~\ref{algo:SynthesisBias}, Line \ref{line:ImprovedWilson}). For the folded-cascode op-amp (Fig. \ref{fig:foldedCascodeOpAmp}), no Wilson voltage bias is created as there is no Wilson current bias in the circuit.  

\begin{algorithm} [tbp]

	\caption{ Creating additional voltage biases} \label{algo:createVoltageBias}
	\begin{algorithmic}[1]
		\REQUIRE  Set of transistors without voltage supply sorted according their doping and position in the op-amp core $T_{un,1, \Phi}$, $T_{un,2, \Phi}$
		\STATE $VB_{\Phi}$ := $\{ ~~\}$ //The set of voltage biases of doping $\Phi$
		\IF{$T_{un,1, \Phi} \cup T_{un, 2, \Phi} = \{cb_{1, n_T = 2}, cb_{2, n_T = 2}, ...\}$} \label{line:CheckingForTwoTransistorCurrentBias}
		\STATE $vb_{\Phi}$ := createCascodeVoltageBias($T_{un,1, \Phi}$, $P_{un,2, \Phi}$) \label{line:createTwoTransistorCurrentBias}
		\STATE $VB_{\Phi}$ := $VB_{\Phi_1} \cup \{vb_{\Phi}\}$
		\ELSE
		\STATE $vb_{1,\Phi}$ :=  createSimpleVoltageBias($T_{un,1, \Phi}$) \label{line:createOneTransistorVoltageBias1}
		\STATE $vb_{2,\Phi}$ :=  createSimpleVoltageBias($T_{un,2, \Phi}$) \label{line:createOneTransistorVoltageBias2}
		\STATE $VB_{\Phi}$ := $VB_{\Phi} \cup \{vb_{1,\Phi}\} \cup \{vb_{2,\Phi}\}$
		\ENDIF
		\RETURN  $VB_\Phi$ 
	\end{algorithmic}
\end{algorithm}

\begin{algorithm} [t] 
	\caption{ Adding current biases} \label{algo:createCurrentBias}
	\begin{algorithmic}[1]
		\REQUIRE Set of voltage biases being of one doping with the bias pin included $VB_{\Phi_{Bias}}$, set of voltage biases being of the other doping $VB_{\Phi_{Other}}$
		\STATE $CB$ := $\{ ~~\}$ //The set of current biases 
		\STATE $VB_{Dis}$ := $\{ ~~\}$ //Set od distributor voltage bias
		\IF{$|VB_{\Phi_{Bias}}| > 1$} 
		\STATE $vb_{Dis}$ := findDistributorVoltageBias($VB_{\Phi_{Other}}$) \label{line:DistributorVoltageBias}
		\STATE $CB_{\Phi_{Other}}$ := createCurrentBiases($vb_{Dis}$, $VB_{\Phi_{Bias}}$)
		\STATE $CB$ := $CB \cup CB_{\Phi_{Other}}$
		\IF{$VB_{\Phi_{Other}}\cap vb_{Dis} = \{~~~\}$}
		\STATE $VB_{Dis}$ := $VB_{Dis} \cup vb_{Dis}$
		\ENDIF
		\ENDIF
		\IF{$VB_{\Phi_{Other}} \neq \emptyset$}
		\STATE $CB_{\Phi_{Bias}}$ := createCurrentBiases($vb_{Bias}$, $VB_{\Phi_{Other}}$)
		\STATE $CB$ := $CB \cup CB_{\Phi_{Bias}}$
		\ENDIF
		\RETURN  $CB, VB_{Dis}$ 
	\end{algorithmic}
\end{algorithm}

To supply the remaining cascode and simple current biases, voltage biases are created with Alg. \ref{algo:createVoltageBias}. If all remaining transistors of doping $\Phi$ in $T_{in, \Phi}$ are part of cascode current biases, a cascode voltage bias is created and connected to supply these transistors (Line \ref{line:createTwoTransistorCurrentBias}). Otherwise, simple voltage biases are created to connect the transistors in the two sets $T_{un,1, \Phi}$, $T_{un,2, \Phi}$ (Line \ref{line:createOneTransistorVoltageBias1}, \ref{line:createOneTransistorVoltageBias2}). If one of the sets is empty, the corresponding voltage bias is not created.
For the folded-cascode op-amp (Fig.~\ref{fig:foldedCascodeOpAmp}), an three voltage biases are created:  $P_{14}$ for $T_{un, p, 1}$, $P_{17}$ for $T_{un, p, 2}$, $N_7$ for $T_{un, n, 2}$.  Also $P_3, P_5$ and $P_4, P_6$ form cascode current biases, as $T_{un, p, 1}$ contains additional transistors not being part of a cascode current bias ($P_7, P_{12}, P_{13}$), no cascode voltage bias is created. 

From the created voltage biases, one voltage bias is chosen to be connected to the current bias input pin $p_{Bias}$ (Alg.~\ref{algo:SynthesisBias}, Line \ref{line:BiasPin}). In the selection process, a single voltage bias is preferred to a cascode one, which is preferred to an improved Wilson  voltage bias.  
This voltage bias $vb_{Bias}$ is already set to be a distributor voltage bias for the later created current biases of the same doping (Fig. \ref{fig:SchematicBias}).
For the folded-cascode op-amp (Fig.~\ref{fig:foldedCascodeOpAmp}), $P_{14}$ is chosen as $vb_{Bias}$. Different to $P_{17}, N_7$, it has a source connection to the supply-voltage rail.

Alg. \ref{algo:createCurrentBias} creates the current biases of the op-amp bias $b_O$. Its input are the sets of voltage biases ordered according to their doping $VB_{\Phi_{Bias}}, VB_{\Phi_{Other}}$.
The algorithm first creates all current biases of doping $\Phi_{Other}$. Current biases of this doping are only needed when the number of voltage biases having the doping $\Phi_{Bias}$ is larger than one. As the current biases must be connected to a {\em distributor voltage bias}, a respective voltage bias must be selected in $VB_{\Phi_{Other}}$ using the same criteria as for $vb_{Bias}$.
If $VB_{\Phi_{Other}}$ is empty or contains only voltage biases supplying transistors in $T_{un,2}$ (Fig. \ref{fig:foldedCascodeOpAmp}, $N_{7}$), a simple voltage bias is created to be the distributor voltage bias $vb_{Dis}$ (Fig. \ref{fig:foldedCascodeOpAmp}, $N_{9}$). 
If voltage biases of doping $\Phi_{Other}$ were created during the synthesis process, current biases of doping $\Phi_{Bias}$ are created. 
The distributor voltage bias for them is $vb_{Bias}$.
For the folded-cascode op-amp (Fig. \ref{fig:foldedCascodeOpAmp}), $VB_{\Phi_{Bias}} = \{P_{14}, P_{17}\}$, $VB_{\Phi_{Other}} = \{N_7\}$.  $P_{14}$ is connected to the bias pin. To bias $P_{17}$, the current bias $N_8$ is created and added to $VB_{\Phi_{Other}}$. As $VB_{\Phi_{Other}}$ does not have a voltage bias connected to the supply-voltage rail, $P_9$ is created as distributor voltage bias of $\Phi_{Other}$. Two current biases $P_{15}, P_{16}$ are created to bias $N_7, N_9 \in VB_{\Phi_{Other}}$.

\section{\underline{FU}nctional \underline{B}l\underline{O}ck \underline{CO}mposition (FUBOCO) of a Complete Op-Amp Topology}\label{sec:synthesisSimpleOpAmp}

\begin{figure*} [tp] \centering
{

	\includegraphics[width=  \linewidth]{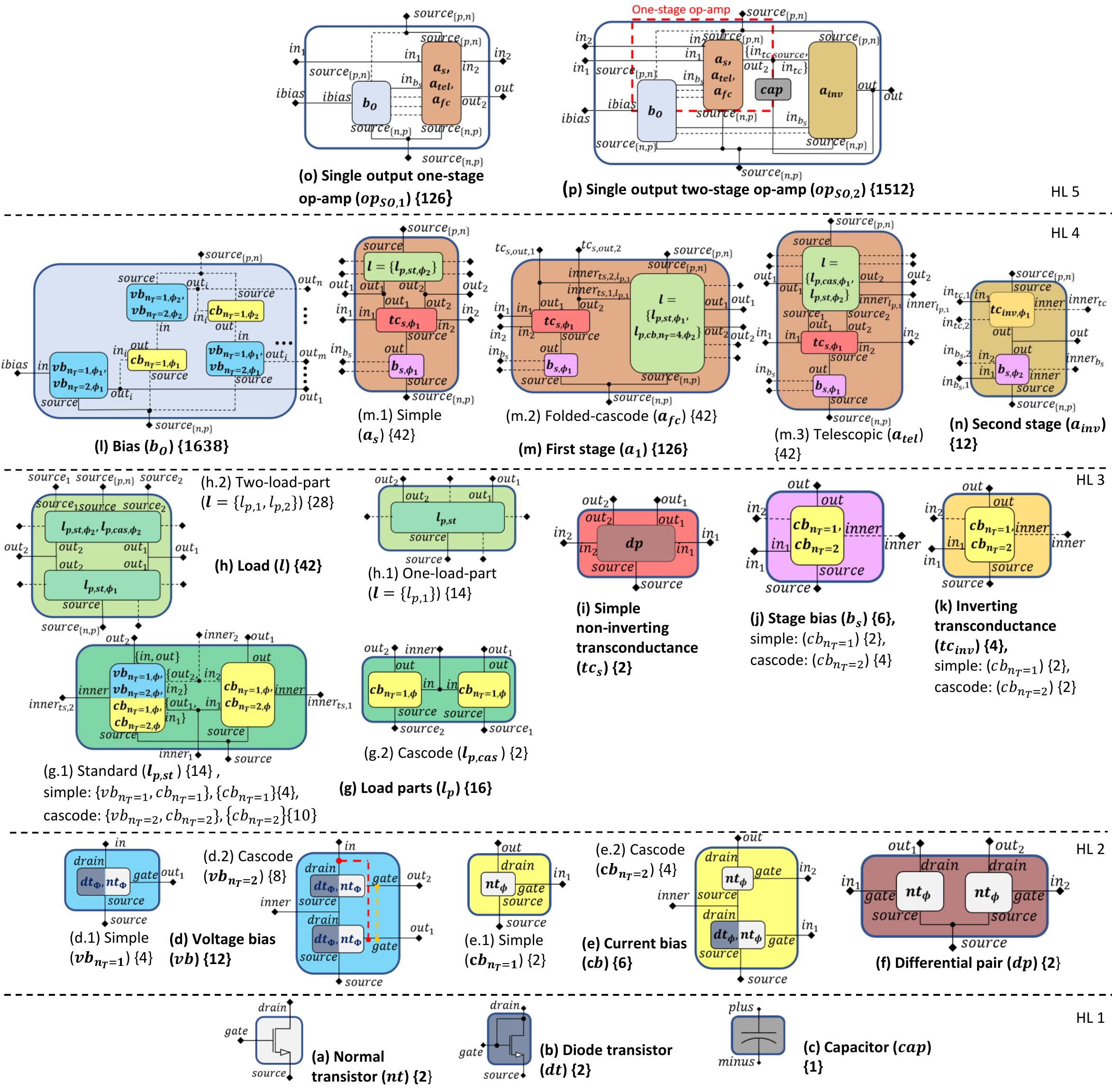}
	\caption{Functional block composition rules for single-output op-amps. $\{n\}$ denotes the respective number of synthesizable structural implementations}
	\label{fig:functionalBlockLibrary}
}
\end{figure*}

The described algorithms (Algs.~\ref{algo:SynthesisFunctionalBlock} - \ref{algo:createCurrentBias}) are combined to synthesize complete op-amp topologies. Fig. \ref{fig:functionalBlockLibrary} shows the composition rules to synthesize all implementations of functional blocks for single-output one-stage and two-stage op-amps. The composition rules for additional functional blocks for single-output symmetrical op-amps (e.g Fig. \ref{fig:symmetricalOpAmp}), fully-differential op-amps (e.g. Fig.~\ref{fig:foldedCascodeOpAmp}) and complementary op-amps (e.g. Fig.~\ref{fig:railToRailAmplifier}) are given in the Appendix. Structural examples synthesizable  for each functional block type are given in \cite{ABNG20}. An outlook on three-stage op-amp synthesis is given in Sec. \ref{sec:OutlookThreeStageOpAmp}.

\subsection{Hierarchy Level 1: Devices}\label{sec:devicesAlgorithmus}
For every device type, instances are created ({Fig~\ref{fig:functionalBlockLibrary}}a-c). 
For transistors, it is differentiated between n- and p-doping ($\Phi_n, \Phi_p$).

\subsection{Hierarchy Level 2: Structures}\label{sec:structureAlgorithmus}
Two types of {\em voltage bias ($vb$)} implementations ({Fig.~\ref{fig:functionalBlockLibrary}}d) are synthesized by Alg. \ref{algo:SynthesisFunctionalBlock}.
For the {\em simple voltage bias} ($vb_{n_T = 1}$, {Fig.~\ref{fig:functionalBlockLibrary}}d.1), only one set implementations is inputted in the algorithm consisting of normal and diode transistors (${S}_1: NT, DT$).
The synthesis of  {\em cascode voltage bias} implementations ($vb_{n_T = 2}$, {Fig.~\ref{fig:functionalBlockLibrary}}d.2) is discussed in Sec. \ref{sec:functionalBlockSynthesis}. 

{\em Current biases}  ($cb$, {Fig~\ref{fig:functionalBlockLibrary}}e) are synthesized similar to the voltage biases. {\em Simple current biases} ($cb_{n_T = 1}$, {Fig.~\ref{fig:functionalBlockLibrary}}e.1) only consist of normal transistors (${S}_1: NT$, e.g. Fig. \ref{fig:foldedCascodeOpAmp} $P_{15}$). The {\em cascode variant} ($cb_{n_T = 2}$, {Fig.~\ref{fig:functionalBlockLibrary}}e.2) consists either of a diode and a normal transistor (e.g. Fig. \ref{fig:railToRailAmplifier} $P_5, P_7$) or two normal transistors (e.g. Fig. \ref{fig:symmetricalOpAmp} $P_7, P_8$) having a drain-source connection (${S}_1: NT_\Phi, DT_{\Phi};$ ${S}_2: NT_\Phi;$ ${R}_c: s_1.drain \leftrightarrow s_2.source$).

{\em Differential pairs} ($dp$, {Fig~\ref{fig:functionalBlockLibrary}}f) are created using two normal transistors ($nt$) of the same doping $\Phi$ connected at the sources (e.g., Fig. \ref{fig:railToRailAmplifier}, $N_1, N_2$) (${S}_1: NT_\Phi;$  ${S}_2: NT_\Phi;$ ${R}_c: s_1.source \leftrightarrow s_2.source;$).

\subsection{Hierarchy Level 3: Amplification Stage Subblocks}\label{sec:amplificationStageSubblocksAlgorithmus}
Two different types of {\em load parts ($l_p$)} are synthesized for single-output op-amps ({Fig.~\ref{fig:functionalBlockLibrary}}g):

The {\em standard load part } ($l_{p,st}$, ({Fig~\ref{fig:functionalBlockLibrary}}g.1) is synthesized based on either of two current biases of the same implementation (e.g. Fig. \ref{fig:railToRailAmplifier} $N_4 - N_7$), or of a voltage and a current bias (e.g. Fig. \ref{fig:railToRailAmplifier} $P_4 - P_7$). The voltage and current biases are either simple (e.g. Fig.~\ref{fig:threeStageOpAmp} $P_1 - P_2$) or cascode (e.g. Fig.~\ref{fig:foldedCascodeOpAmp} $P_3 - P_6$). The functional subblocks $s_1, s_2$ are connected at their sources. Also the inputs are connected or, in cases of a load part based on a voltage and a current bias, the outputs are connected to the inputs.
\begin{lstlisting}[mathescape=true]
${S}_1: CB_{\Phi}, VB_{\Phi};$  ${S}_2: CB_\Phi;$ 
${R}_c: n_{T,fb_1} = n_{T, fb_2},$$s_1.source$$\leftrightarrow s_2.source, (s_1.in_1 \vee s_1.out_1) \leftrightarrow s_2.in_1,$$\forall_{n_{T, s_1} = 2} (s_1.in_2 \vee s_1.out_2) \leftrightarrow s_2.in_2;$ 
\end{lstlisting}

{\em Cascode load part } implementations ($l_{p, cas}$, Fig. {\ref{fig:functionalBlockLibrary}g.2) are only relevant for the synthesis of telescopic op-amps.  The load part is synthesized based on simple current biases ($cb$) being only connected at the input pins (${S}_1: CB_{n_T = 1,\Phi};$  ${S}_2: CB_{n_T = 1,\Phi};$ ${R}_c: s_1.in_1 \leftrightarrow s_2.in_1;$).

{\em Loads } ($l$, Fig. {\ref{fig:functionalBlockLibrary}}h) are synthesized using either one (Fig.~\ref{fig:symmetricalOpAmp}, $P_1 -P_4$) or two load parts (Fig. \ref{fig:foldedCascodeOpAmp}, $P_3 - P_6, N_1 - N_4$).

{\em One-load-part loads} implementations ($l= \{l_{p,1}\}$, Fig. {\ref{fig:functionalBlockLibrary}}h.1) are created with standard load parts ($S_1: L_{p,st}$).

{\em Two-load-part loads} ($l= \{l_{p,1}, l_{p,2}\}$, Fig. {\ref{fig:functionalBlockLibrary}}h.2) consist of two standard load parts $l_{p, st, \Phi_1}$, $l_{p, st, \Phi_2}$ with different doping (e.g. Fig. \ref{fig:foldedCascodeOpAmp} $N_1 - N_4$, $P_3 - P_6$) or, iff the op-amp has a telescopic first stage, a standard load part $l_{p, st, \Phi_1}$ and a cascode load parts $l_{p, cas, \Phi_2}$.
They are connected at the load part outputs.
(${S}_1:  L_{p, st, \Phi_1};$  ${S}_2: L_{p, st, \Phi_2},$$ L_{p,cas, \Phi_2};$ 
${R}_c: s_1.out_1 \leftrightarrow s_2.out_1, s_1.out_2 \leftrightarrow s_2.out_2;$ )

{\em Simple non-inverting transconductances}  ($tc_{s}$, Fig. {\ref{fig:functionalBlockLibrary}}i) are synthesized based on one differential pair (${S}_1: DP$) (e.g. Fig.~\ref{fig:foldedCascodeOpAmp} $P_1, P_2$).

{\em Stage biases } ($b_s$, Fig. {\ref{fig:functionalBlockLibrary}}j) are created  based on simple (Fig.~\ref{fig:foldedCascodeOpAmp} $P_7$) or cascode current biases (${S}_1:  CB$).

{\em Inverting transconductances} ($tc_{inv}$, Fig. \ref{fig:functionalBlockLibrary}k) are based on  current biases $CB$ (${S}_1:  CB;$), which are simple   or cascode. No connection between the first input pin and the inner pin is allowed (${\mathcal{R}}_f: s_1.in_1 \nleftrightarrow s_1.inner$).

\subsection{Hierarchy Level 4: Amplification Stages}\label{sec:opAmpSubblocksAlgorithmus}
The topology-specific {\em op-amp bias $b_O$} (Fig. \ref{fig:functionalBlockLibrary}l) is synthesized  using Alg. \ref{algo:SynthesisBias} after the amplification stages and capacaitors are created and connected. The bias consists of voltage and current biases (e.g. Fig. \ref{fig:foldedCascodeOpAmp} $P_{14} - P_{17}$, $N_7 -N_9$).

Three different types of {\em first stage} implementations ($a_{1}$,  Fig.~{\ref{fig:functionalBlockLibrary}}m) are supported for simple op-amps:
 
{\em Simple first stages} ($a_s$, Fig. {\ref{fig:functionalBlockLibrary}}m.1) are synthesized based on a one-load-part load, a simple non-inverting transconductance and a stage bias (${S}_1:  TC_{s, \Phi_1};$  ${S}_2: B_{s, \Phi_1};{S}_3: L = \{L_{p,st}\}$). The load is of different doping $\Phi_2$ than the transconductance and stage bias $(\Phi_1$).
The transconductance's source is connected to the output of the stage bias, while its outputs are connected to the outputs of the load (${R}_c: s_1.source$  $\leftrightarrow s_2.out,$ $s_1.out_1 \leftrightarrow s_3.out_1,$$s_1.out_2 \leftrightarrow s_3.out_2;$).

{\em Folded-cascode first stage}  implementations ($a_{fc}$, Fig. {\ref{fig:functionalBlockLibrary}}m.2) are  synthesized  with loads consisting of two standard load parts. One  load part $l_{p,cb, n_T = 4, \Phi_2}$ of the two-load-part load consists of current biases, has four transistors and  a different doping than $tc_s$. This load part is connected with its inner pins of the current biases $inner_{ts_1, l_{p,1}}, inner_{ts_2, l_{p,1}}$ to the output pins of the transconductance. %The other load part consists of a voltage and a current bias. 
\begin{lstlisting}[mathescape=true]
${S}_1:  TC_{s, \Phi_1};$  ${S}_2: B_{s, \Phi_1};$ ${S}_3: L = \{L_{p,st, \Phi_1},L_{p,cb, n_T = 4, \Phi_2}\};$ 
${R}_c: s_1.source \leftrightarrow s_2.out,$ $s_1.out_1 \leftrightarrow s_3.inner_{ts_1, l_{p,1}},$$s_1.out_2 \leftrightarrow s_3.inner_{ts_2, l_{p,1}};$ 
\end{lstlisting}

{\em Telescopic first stages} ($a_{tel}$, Fig. {\ref{fig:functionalBlockLibrary}}m.3) are created with a two-load part load consisting of a cascode  and a standard load part $ L = \{L_{p,cas, \Phi_1},L_{p,st, \Phi_2}\}$ 
\begin{lstlisting}[mathescape=true]
${S}_1:  TC_{s, \Phi_1};$  ${S}_2: B_{s, \Phi_1};$ ${S}_3: L = \{L_{p,cas, \Phi_1},$$L_{p,st, \Phi_2}\};$ 
${R}_c: s_1.source \leftrightarrow s_2.out,$$s_1.out_1 \leftrightarrow s_3.source_1,$$s_1.out_2 \leftrightarrow s_3.source_2;$ 
\end{lstlisting}

{\em Second stages} ($a_{inv}$, Fig. {\ref{fig:functionalBlockLibrary}}n) are composed of an inverting transconductance $tc_{inv}$ and a stage bias $b_s$ of different doping (${S}_1:  TC_{inv, \Phi_1};$  ${S}_2: B_{s, \Phi_2}$). They are connected at their outputs
(${R}_c: s_1.out \leftrightarrow s_2.out,$).

\subsection{Hierarchy Level 5: Op-Amps}\label{sec:opAmpsSynthesis}
Two types of {\em single output op-amps} ($op_{SO}$) are supported:

{\em One-stage op-amp}  implementations ($op_{SO,1}$, Fig. {\ref{fig:functionalBlockLibrary}}n) are created by synthesizing different first stage implementations (${S}_1:  A_{s}, A_{tel},  A_{fc}$) with Alg. \ref{algo:SynthesisFunctionalBlock}. The bias circuit $b_O$ is synthesized with Alg. \ref{algo:SynthesisBias}.

{\em Two-stage op-amps} ($op_{SO,2}$, Fig. {\ref{fig:functionalBlockLibrary}}p) are created  by adding a second stage $a_{inv}$ and a capacitor $cap$ to the first stage of the one-stage op-amp  using Alg. \ref{algo:SynthesisFunctionalBlock} (${S}_1:  A_{s}, A_{tel}, A_{fc};$  ${S}_2: cap;$ ${S}_3: A_{inv};$). The capacitor is connected between the output of the first stage and the second stage, the output of the first stage is connected to the input of the second stage (${R}_c:s_1.out_2$
$\leftrightarrow s_2.plus,$$s_1.out_2 \leftrightarrow s_3.in_{tc,1},$$s_2.minus \leftrightarrow s_3.out;$).
The bias $b_O$ is synthesized with Alg. \ref{algo:SynthesisBias}.

\subsection{Number of implementations per functional block}\label{sec:implementationNumber}
Fig. \ref{fig:functionalBlockLibrary} gives the maximum number of structural implementations per functional block currently supported by the synthesis algorithm. It includes n-type as  well as p-type implementations
and can be controlled by adding/removing rules to the set of functional block rules $R_{f}$ and adding/removing functional block implementation from $S_i$. 
E.g., removing the diode transistors from $S_1$ of the cascode current bias  leads to only two implementation of the cascode current bias instead of four, reducing also the number of synthesized stage biases and thus also the number of first and second stages as well as overall op-amp topologies.

\section{Overview of the Complete FUBOCO Synthesis Process}\label{sec:synthesisAlgorithm}

\begin{figure*} [] \centering
	\subfloat[Functional block composition graph (without op-amp bias). HL1 - HL2: basic functional blocks, HL3 - HL4, thick frame: basic functional blocks of a specific op-amp type.]{
		\label{fig:decisionTree}
		\includegraphics[width= 0.65\linewidth]{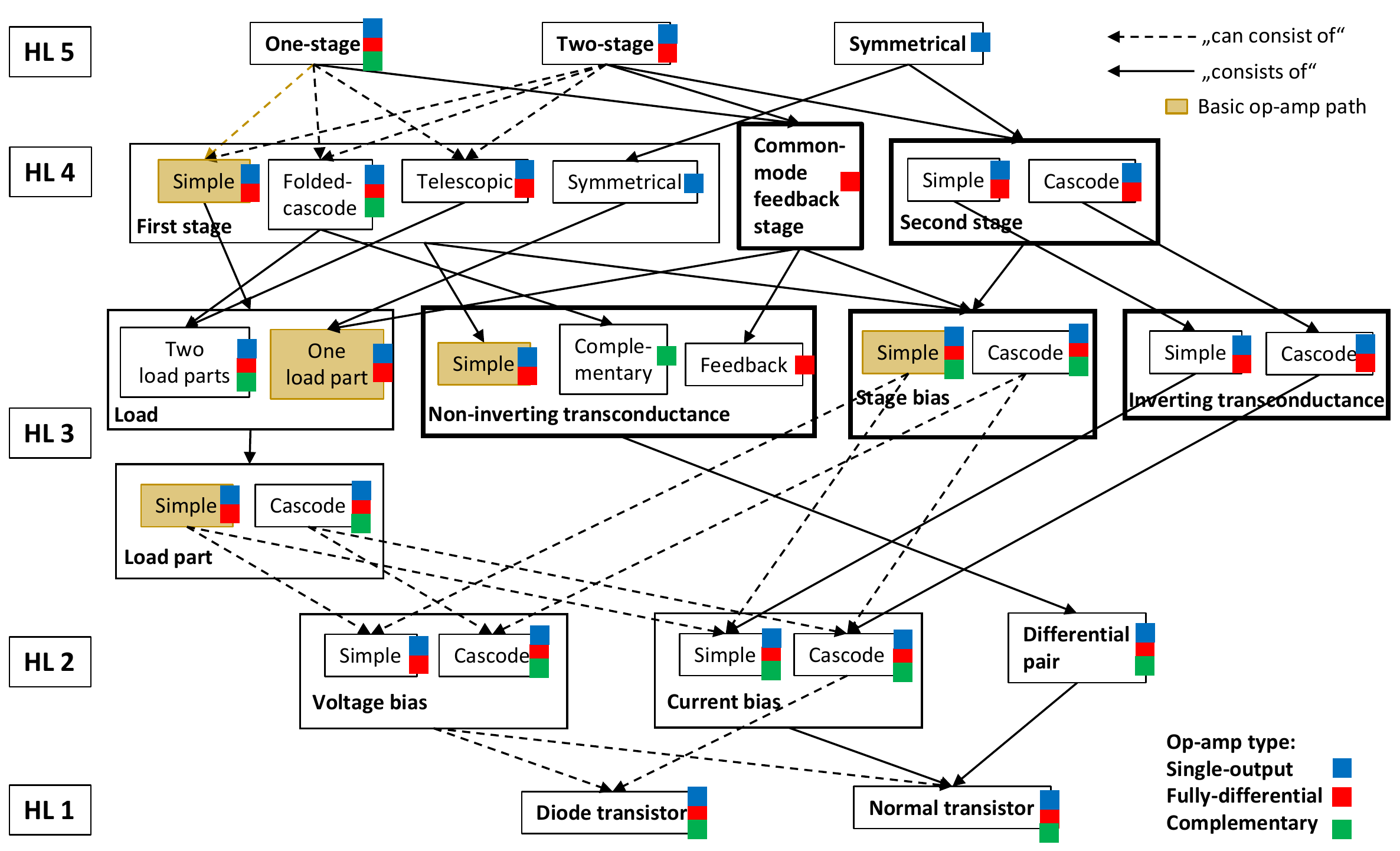}
	}%                           
	\qquad	
	\subfloat[Structure synthesis algorithm.]{
		\label{fig:synthesisAlgorithm}
		\includegraphics[width= 0.28\linewidth]{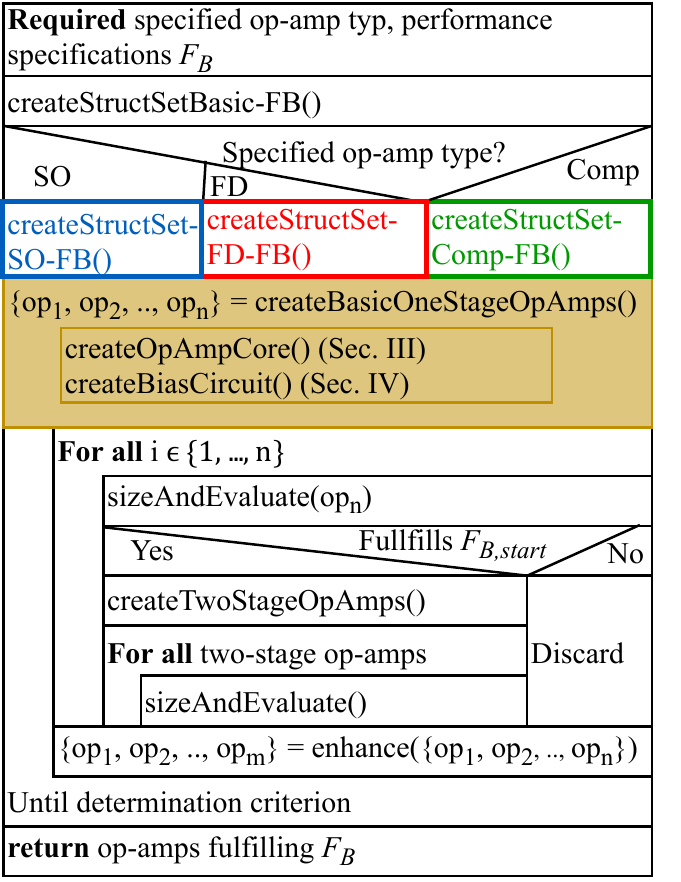}
	}% 
	\qquad
	\caption{Overview of the FUBOCO synthesis process.}
	\label{fig:synthesisTool}
\end{figure*}

Fig.~\ref{fig:synthesisTool} gives an overview of the overall synthesis algorithm. The parts of the hierarchical functional block composition graph (Fig. \ref{fig:decisionTree}) marked with blue dots represent an abstraction of the composition graph for single-output op-amps (SO) in Fig.~\ref{fig:functionalBlockLibrary}. The parts marked with red and green dots represent the functional blocks needed to synthesize  fully-differential (FD, red) and complementary op-amps (Comp, green). Details of the supplementary functional blocks are given in the Appendix.
The op-amp type to be synthesized and its performance requirements $F_B$ are specified by the user.

\subsection{\underline{Fu}nctional \underline{B}l\underline{o}ck \underline{Co}mposition (FUBOCO) Graph }
The composition graph (Fig. \ref{fig:decisionTree}) defines how to compose each functional block on level $x$ from a set of functional blocks on level $x$-$1$ or $x$  by combination, starting from the functional blocks on the 5th hierarchy level, i.e., one-stage, two-stage, or symmetrical op-amp. The usage  of a functional block is either strict (“consists of”), or a selection of one out of many ("can consist of").  
The new composition graph differs from the design plan-based structural synthesis approaches \cite{IntegerBasedTopologieSelecting, AGenericTopologySelectionMethodForAnalogCircuits}, which feature a single solution path through such a composition graph based on if-then-else decisions, and from the structural synthesis approaches that use local structural changes with a nearly open-end process. 
The FUBOCO graph instead provides a large search space of yet only technically meaningful structures and uses a fast equation-based sizing process for an optimization-based selection process.

Each op-amp type features its own part of the functional block composition graph for synthesis. The respective blocks are marked with the respective color in Fig.~\ref{fig:decisionTree}. The common-mode feedback stage (CMFB stage) for instance is only used in
fully-differential op-amps (red), a folded cascode first stage is used for all three types of op-amps (blue, red, green). One-stage, two-stage and symmetrical types of single-output op-amps are considered, one- and two-stage versions of fully-differential op-amps can be synthesized, and one-stage complementary op-amps are supported. An outlook how three-stage op-amps can be integrated in the composition graph is given in Sec. \ref{sec:OutlookThreeStageOpAmp}.

\subsection{Synthesis Algorithm}\label{sec:overallAlgorithm}
The synthesis algorithm, sketched in Fig.~\ref{fig:synthesisAlgorithm}, features a combination of enumerative and generative approach to structure synthesis. The evaluation of created structural op-amp variants is based on optimization over behavorial equations and is particularly fast.
 
Three groups of functional blocks are distinguished and specifically treated in the structural synthesis process:

{\em Basic functional blocks} are all functional blocks of HL~1~- HL 2 (Fig. \ref{fig:decisionTree}). They are part of many other functional blocks, e.g. non-inverting and inverting transconductance, load and stage bias.  The number of implementations per functional block is small (2 - 12 for each type of functional block).

{\em Basic functional blocks of a specific op-amp type} are the functional blocks of HL 3 - HL 4 framed with thick lines. Not all functional blocks are part of every op-amp type, such that the functional blocks can be further divided into op-amp type specific groups having overlaps. The stage bias for instance is part of every group, the common-mode feedback stage is only relevant for fully-differential op-amps. The number of implementations per  functional blocks is small (2 - 12).

{\em Functional blocks with many implementations} are the op-amps themselves, the first stages, loads and load parts. The number of implementations per functional block is high and varies between 24 for the load parts and 318 for the first stages. 

Based on the methods presented in the preceding sections, the algorithm creates all implementations of basic functional blocks and of op-amp type-specific functional blocks upfront and stores them in a library (parts marked with blue, red, green dots in Fig.~\ref{fig:synthesisAlgorithm}). 
Functional blocks with a large number of structural implementation are only created on-demand when they are part of a topology. 
This provides a good comprise between computation time and memory usage.

The algorithm creates a set of basic one-stage op-amps with a low number of transistors. This set is marked with golden background in
Fig.~\ref{fig:decisionTree} and refers to the part similarly colored in the synthesis algorithm in Fig.~\ref{fig:synthesisAlgorithm}.
The topologies are evaluated based on sizing (Sec. \ref{sec:evaluation}).
Some op-amp characteristics $F_{start}$  degenerate or do not change if a second stage is added to a one-stage op-amp. If a one-stage op-amp does not fulfill the corresponding specifications $F_{B, start}$, its two-stage versions would also not fulfill the specifications. Hence, two-stage variants are only created if these performance specifications are satisfied.
Otherwise, a one-stage op-amp variant and its potential two-stage variants are discarded, thus bounding and reducing the search tree from irrelevant branches.

One-stage op-amps are enhanced by changing their stage bias and/or load. A new set of one-stage op-amps  and their two-stage variants is configured and evaluated.
The synthesis process ends if either the simplest op-amp fulfilling the specification is found or all op-amp topologies are enumerated.

Please note that complementary op-amps and symmetrical op-amps are synthesized slightly different. For complementary op-amps, only one-stage op-amps are currently considered.
Symmetrical op-amps exist only as two-stage variants, such that the differentiation between one- and two-stage variant is not made.

\subsection{Topology Sizing and Evaluation}\label{sec:evaluation}

Topologies are evaluated with an equation-based sizing method. Based on standard equations, as e.g. in \cite{LakerSansen}, a behavioral model based on analytical equations for each functional block in Fig. \ref{fig:FunctionalBlockLibrary} was developed leading to a hierarchical performance equation library \cite{ABNG20c}.
The open-loop gain $A_{D,0}$ for examples is calculated by the multiplication of the open-loop gain of the op-amp stages $n$:
\begin{equation}
	A_{D,0} = \prod_{i=1}^n A_{D,i}; ~~~ A_{D, i} = gm_i \cdot R_{out}
\end{equation} 
The open-loop gain of an amplification stage $A_{D, i}$  is calculated by the input transconductance of the stage $gm_i$ and its output resistance $R_{out}$. As the structure of the op-amp and its amplification stages is known through the composition graph (Fig. \ref{fig:functionalBlockLibrary}), the equations for an op-amps are automatically instantiated as described in \cite{ABNG20c}.

The optimization approach within initial sizing method is an enhanced version of~\cite{ABNG20b}.
Two groups of performance variables are defined.
The group $F_{start}$ contains all performance variables which depend only on design variables of the first stage or degrade by adding a second stage to the one-stage op-amp. These are the transistor gate-area $f_{D}$, the power consumption $f_{P}$, the maximum and minimum common-mode input voltage $f_{v_{cm,max}}, f_{v_{cm,min}}$,  the common-mode rejection ratio $f_{CMRR}$ and the phase margin $f_{PM}$.
\begin{equation}\label{eq:Fstart}
\begin{split}
F_{start} = \{&f_{D}, f_{P}, f_{v_{cm,max}}, f_{v_{cm,min}}, f_{CMRR}, f_{PM} \}
\end{split}
\end{equation}
The group $F_{end}$ contains the remaining performance variables. These are the open-loop gain $f_{A_{D,0}}$, the slew rate $f_{SR}$, the unity-gain bandwidth $f_{GBW}$ and the maximum and minimum output voltage swing $f_{v_{out,max}}, f_{v_{out,min}}$:
\begin{equation}\label{eq:FBend}
F_{end} = \{f_{A_{D,0}}, f_{SR}, f_{GBW}, f_{v_{out,max}}, f_{v_{out,min}}\}
\end{equation}

The performance variables in the set $F_{start}$ must  fulfill the user-given constraint values $F_{B,start}$. 
Otherwise the topology is discarded from the synthesis process. If it was a one-stage op-amp, its two-stage variants are not  considered in the synthesis process.

The variables of $F_{end}$ are  optimized towards over-fulfillment of their specifications $F_{B,end}$.
If this succeeds, all performance variables $F = F_{start} \cup F_{end}$ are further optimized. 
In this work, the optimization algorithm is based on constraint programming and does not seek to reach an optimum but is terminated when the optimization progress slows down. Experiments have shown that after an optimization time of around one minute per structural variant the progress slows down. We therefore set the time limit for one optimization run to one minute.

The analytical equation-based sizing method allows the emulation of the manual sizing process during topology evaluation, this is different to other approaches, e.g. \cite{FEATS, VariationAwareStructuralSynthesisOfAnalogCircuitsViaHierarchicalBuildingBlocksAndStructuralHomotopy}, using numerical, simulation-based sizing methods \cite{ANGW94,OCRC96, Reviewer1}.

\section{Experimental Results}\label{sec:experimentalResults}
We tested the synthesis tool with seven different specification sets (Table \ref{tab:Specifications}). FUBOCO creates  for each set all topologies that fulfill the specifications. The user specifies the op-amp type (single-output, fully-differential, complementary) and the performance requirements in the specification set. Additional design knowledge is not required. 
All supported  implementations of every functional block were allowed in the synthesis process (Sec.~\ref{sec:implementationNumber}).

{\em Specs 1, Specs 2} and {\em Specs~3} (Table \ref{tab:Specifications}) specify single-output as  op-amp type.  {\em Specs 1} has a high gain requirement likely to exclude many topologies.
{\em Specs 2}  demands a smaller quiescent power and gate-area. Also the  requirements for slew rate and unity-gain bandwidth are more challenging.
{\em Specs~3} requires an even smaller quiescent power. The requirements for the gain are less strict. A high phase margin is required.
{\em Specs 4} and {\em Specs 5} are specified  for fully-differential op-amps. In {\em Specs 4}, the most demanding specification is the gain being comparatively high. 
{\em Specs 5}  demands a smaller area and quiescent power and a higher unity gain-bandwidth and slew rate.
The op-amp type  with {\em Specs 6} and {\em Specs 7} is complementary. {\em Specs 6} are satisfiable by many topologies variants of complementary op-amps.
In {\em Specs 7}, the allowed area and quiescent power is reduced. The  requirements for unity-gain bandwidth and slew rate increase.

Please note that only the gate-area of every transistor $t \in T$, i.e. $\sum_{t \in T} W_t L_t$, is considered as area constraint. However, also other models including more layout aspects can be used in the calculation.

\begin{table}[tbp]
	\centering	\setlength{\tabcolsep}{0.1cm}	
	\caption{Specifications; {Specs} 1, Specs 4, Specs 5: general Specs; Specs 2, Specs 3, Specs 5, Specs 7: sophisticated Specs}\label{tab:Specifications}
	\begin{tabular}{|l||c|c|c||c|c||c|c|}
		\hline	
		{\em Specs} \# &  1 &  2&  3 &  4 &  5	& 6 &  7\\ \hline\hline
		Op-amp type &\multicolumn{3}{c||}{ Single-output} & \multicolumn{2}{c||}{Fully-differential} & \multicolumn{2}{c|}{Complementary} \\ \hline
	 
		Bias current ($\mu$A) & 10 & 100 & 10 & 100 & 100 & 100 & 100 \\
		Load Capacity (pF) & 20 & 20 & 20 & 20 & 20 & 20 & 20 \\
		Supply voltage (V) & 5 & 5 & 5 & 5& 5& 5& 5\\
		Gate-area ($10^3 \mu$m$^2$) & $\leq$ 15& $\leq$ 5& $\leq$ 5 & $\leq$ 50 & $\leq$ 20& $\leq$ 15  & $\leq$ 5\\\hline
		Quiescent power (mW) & $\leq$ 15 &  $\leq$ 8&  $\leq$ 5 &  $\leq$ 25 &  $\leq$ 15 &  $\leq$ 10 &  $\leq$ 5 \\
		Phase Margin ($^\circ$) & $\geq$ 60  & $\geq$ 60 & $\geq$ 80 & $\geq$ 60 & $\geq$ 60 & $\geq$ 60 & $\geq$ 60\\
		CMRR (dB) & $\geq$ 70& $\geq$  70  & $\geq$  70 & $\geq$  80 & $\geq$  80 & $\geq$  80 & $\geq$  80\\
		CMIR (V) & 2-3 &  1.5-3.5 & 1.5-3.5 & 2-3 & 2-3 & -&-\\
		Open-loop gain (dB) & $\geq$ 80 & $\geq$ 70 & $\geq$ 45 & $\geq$ 70 & $\geq$ 60 & $\geq$ 70 & $\geq$ 70\\
		Unity-gain bandwidth (MHz) &$\geq$ 2.5 &$\geq$ 10 &$\geq$ 10 &$\geq$ 2.5 &$\geq$ 10 &$\geq$ 2.5 &$\geq$ 10\\
		Slew rate ($\frac{\text{V}}{\mu \text{s}}$)& $\geq$ 3.5& $\geq$ 20 & $\geq$ 20 & $\geq$ 3.5 & $\geq$ 15 & $\geq$ 3.5 & $\geq$ 15\\
		Output voltage swing (V) & 1.5-3.5& 1.5-3.5 & 1-4 & 2-3 & 1.5-3.5 & 1.5-3.5 & 1-4 \\\hline
	\end{tabular}
\end{table}

\subsection{Synthesized Topologies}\label{sec:structuralSynthesis}

Table \ref{tab:PerformanceParameters} shows the number of topologies created by FUBOCO for the different specifications.  FUBOCO currently supports 2940 single-output topologies of which 210 are one-stage topologies, 936 fully-differential topologies (72 one-stage/864 two-stage) and 36 complementary op-amps.
Not all topologies are created in each run, as the syntheses process does not consider two-stage op-amps if its one-stage variant fail the specifications in $F_{B,start}$ (Sec. \ref{sec:overallAlgorithm}). 

Table \ref{tab:structureResults} gives an overview of the composition of amplification stages in the synthesized topologies  for {\em Specs~1} - {\em Specs~5}. Topologies with a large structural variety are outputted for general specifications ({\em Specs~1}, {\em Specs~4}), a smaller variety results for more specific specifications  ({\em Specs~2}, {\em Specs~3}, {\em Specs~5}). 

For {\em Specs~1}, the largest number of op-amps were created. Many one-stage op-amps fullfilled $F_{B,start}$ (144), such that in total 1728 op-amps were created.  228 of the 1728 topologies fulfilled {\em Specs~1}.
Due to the high gain requirement, most of topologies are two-stage op-amps with a simple or a folded-cascode first stage (Table \ref{tab:structureResults}). The one-stage op-amps are either formed with a folded-cascode or telescopic first stage.
The 60 symmetrical op-amps have all a cascode second stage. 
An example of an op-amp fulfilling the specifications is the symmetrical op-amp in Fig. \ref{fig:symmetricalOpAmp}. 

As {\em Specs 2} is more strict, the number of topologies fulfilling the set  is much smaller: 96 one-stage op-amps fulfill $F_{B,start}$ leading to 1152 created op-amps in total. 92 of the 1152 topologies fulfill all specifications.
The set is dominated by telescopic and folded-cascode one-stage op-amps and symmetrical op-amps (Table \ref{tab:structureResults}). Due to the strong area constraint, folded-cascode two-stage op-amps do not longer fulfill the specifications. The small number of two-stage op-amps have all a simple first stage.
As the constraints on the input voltage are more demanding, all topologies in the set have a simple voltage bias as stage bias of the first stage. 
The symmetrical op-amp (Fig. \ref{fig:symmetricalOpAmp}) is also a valid topology for {\em Specs~2}.

\begin{table}[t]
\centering	\setlength{\tabcolsep}{0.1cm}
	\caption{Number of created topologies by FUBOCO and resulting runtime; brackets: max. \# supported topologies}\label{tab:PerformanceParameters}
	\begin{tabular}{|l||c|c|c||c|c||c|c|}
	\hline
	{\em Specs} \# &  1 &  2&  3 &  4 &  5	& 6 &  7\\ \hline \hline
	\# one-stage op-amps fulfilling $F_{B,start}$ & 144 (210) & 96 (210) & 81 (210) & 39 (72) & 37 (72) & 36 (36) & 36 (36)\\
	\# created op-amp topologies & 1728 (2940) & 1152 (2940) & 972 (2940) & 468 (936) & 444 (936) & 36 (36) & 36 (36)\\
	\# op-amps fulfilling $F_{B}$ & 228 (2940) & 71 (2940) & 54 (2940) & 34 (936) & 2 (936) & 10 (36) & 6 (36)\\\hline
	Runtime & 21 h & 16 h & 14.5 h & 8 h & 8 h & 35 min & 30 min \\\hline
	\end{tabular}
\end{table}

\begin{table}[tbp]
	\centering	\setlength{\tabcolsep}{0.1cm}	
	\caption{Amplification stage composition of the resulting topologies}\label{tab:structureResults}
	\begin{tabular}{|l||>{\centering\arraybackslash}m{0.6cm}|>{\centering\arraybackslash}m{0.5cm}||>{\centering\arraybackslash}m{1.25cm}|>{\centering\arraybackslash}m{1.1cm}||>{\centering\arraybackslash}m{0.9cm}|>{\centering\arraybackslash}m{0.8cm}||c||c|}
		\hline	
		First stage type &\multicolumn{2}{>{\centering\arraybackslash}m{1.5cm}||}{ simple $a_s$} & \multicolumn{2}{>{\centering\arraybackslash}m{2.5cm}||}{folded-cas\-code $a_{fc}$} & \multicolumn{2}{>{\centering\arraybackslash}m{2cm}||}{telescopic $a_{tel}$} & symmetrical $a_{sym}$ & \multirow{2}{2cm}{total \# topologies} \\ \cline{1-8}
		\# stages & 1 & 2 & 1 & 2 & 1 & 2 & - &\\ \hline \hline
		{\em Specs 1} & 0 &	40& 30&68& 30&0&60&	228 \\ \hline
		{\em Specs 2} & 0 & 5 & 18 & 0 & 20 & 0 & 28 & 71 \\ \hline
		{\em Specs 3} & 24 & 0 & 0 & 0 & 3 & 0 & 27 & 54 \\ \hline \hline
		{\em Specs 4} & 0 &	0 &	11 & 22 & 1	& 0 &- & 34\\ \hline
		{\em Specs 5} & 0 &	0 &	2 &	0 &	0 &	0 &	- &	2 \\ \hline
	\end{tabular}
\end{table}

The set of topologies fulfilling {\em Specs 3} is even smaller.  As 81
one-stage op-amps fulfilled the specifications in $F_{B, start}$, 972 op-amps in total were created. 54 topologies fulfilled all specifications $F_{B}$.
Only one-stage op-amps and symmetrical op-amps fulfill the specifications due to the high phase margin constraint (Table \ref{tab:structureResults}). The one-stage op-amps have mostly a simple first stage. The three telescopic op-amps have all a simple current mirror as one of the load parts. The second stages in the symmetrical op-amps are mostly simple. 

For {\em Specs 4}, 468 of the 936 fully-differential op-amp topologies were created in total, as 39 one-stage op-amps fulfilled the specifications in $F_{B,start}$. 34 topologie fulfilled  all specifications.
These are mainly folded-cascode one-stage and two-stage  op-amps (Table \ref{tab:structureResults}). Also one telescopic one-stage op-amp fulfilled the specifications. 
An example of a fully-differential op-amp topology fulfilling {\em Specs 4} is shown in Fig. \ref{fig:foldedCascodeOpAmp}. It is a folded-cascode one-stage op-amp with a pmos differential stage.

The number of topologies fulfilling  {\em Specs 5}  is smallest in this scenario: Similar to  {\em Specs 4}, 37 topologies fulfilled $F_{B,start}$ leading to 444 created op-amps. Only two  folded cascode one-stage topologies fulfilled all specifications (Table~\ref{tab:structureResults}).
One is the topology shown in Fig. \ref{fig:foldedCascodeOpAmp}. The other one is an NMOS version of it. The NMOS version has a cascode stage bias in the first stage instead of a simple one.

As currently only one-stage complementary op-amps are supported by FUBOCO, all 36 topologies are created for every run specified for complementary op-amps.
10 topologies fulfill all specifications in {\em Specs 6}.
They  only vary in their type of loads and stage biases as they all have a folded-cascode first stage. The loads are mainly loads with one load part being a current mirror and the other load part containing two current biases (Fig. \ref{fig:railToRailAmplifier}). Various types of current mirrors as load fulfill the specifications. All types of stage biases appear in the topologies.

Six complementary op-amp topologies fulfill {\em Specs 7}.
They do not differ much from the op-amps fulfilling {\em Specs 6}. The simple stage bias in the first stage dominates the set. The topology in Fig. \ref{fig:railToRailAmplifier} also fulfills {\em Specs 7}.

Please note that topologies might overfill specifications with quite high margin, as only one specification bound is given per performance feature. This can be omitted by using an upper and lower bound for each performance feature, e.g., a maximum and minimum open-loop gain requirements. This would reduce the resulting topologies to only topologies performing very close to the specifications. For specifications with low demands, e.g. {\em Specs 1}, a smaller number of topologies would be outputted.

\subsection{Synthesis runtime}
The Experiments were run on an Intel\textsuperscript{\textregistered} Core\texttrademark ~i5-7500 CPU@3.4GHz with 32 GB RAM. 
The sizing of an op-amp with its optimization time in the range of minutes per circuit is the biggest time constraint of the synthesis tool.  In contrast, the creation of a topology is in the range of milliseconds. 

The runtime highly varies with the number of created op-amp topologies (Table \ref{tab:PerformanceParameters}).
For a single-output op-amp specification set, many topologies can be synthesized and sized ($2940$). For such specifications, the run time of the synthesis tool is quite long (14 h - 22 h). For  complementary op-amps with only 36 topologies supported, the runtime is smaller ($\sim 30$ min).
The requirements of the specifications have a great influence on the runtime. If they are more strict, the runtime decreases for  two reasons:
\begin{itemize}
	\item Strict specifications lessen the number of two-stage op-amps which are created and sized. Many one-stage op-amps do not fulfill the specifications in $F_{B,start}$. Thus, two-stage op-amps based on them are not created.
	\item The constraint-programming solver does not optimize circuits if early results show that the circuit does not fulfill the specifications.
\end{itemize}
As many topologies were created for {\em Specs 1} (1728), 21~h were needed to evaluate them. For {\em Specs 2}, only 1152 topologies were created at a  runtime of around 16~h. For {\em Specs 3}, the number of created topologies was 972, and the overall runtime 14.5 h.
For fully-differential op-amps, the runtime for both specification sets is equal. The total number of one-stage op-amps is smaller compared to single-output op-amps, such that the number of one-stage op-amps fulfilling $F_{start}$ does not vary as much.
For complementary op-amps, only one-stage op-amps are supported such that always 36 op-amps are created and evaluated. For more challenging specs ({\em Specs~7}), the runtime reduces further, as the number of topologies increases for which the sizings are not further optimized as the topologies will not fulfill the specifications.

As the runtime for specifications with many created topologies is quite long, future work remains in reducing the runtime. We are currently working on paralleling the synthesis process, sizing several op-amps at the same time. Depending on the CPU, this reduces the runtime significantly.

\subsection{Sizing and Evaluation Results}\label{sec:sizingAndEvaluationResults}

\begin{table}[tbp]
	\centering	\setlength{\tabcolsep}{0.1cm}
		\caption{Comparison to simulation results; Deviation in per cent; (i/j), i: \# of topologies fulfilling the specs, j. total \# of topologies tested}\label{tab:SizingResults}
	\begin{tabular}{|l||c|c|c||c|c||c|c|}
		\hline
	
		{\em Specs} \# &  1 &  2&  3 &  4 &  5	& 6 &  7\\  \hline	\hline
		Quiescent power  &8\% (25/30) & 7\% (28/28) & 6\%  (21/23) & 4\% (10/10) &  4\% (0/2) &20\% (9/10) & 6\% (6/6)\\ 
		Phase Margin &24\% (25/30) & 11\% (28/28) & 6\%  (16/23) & 3\% (10/10) &  3\% (2/2) &16\% (10/10) & 15\% (5/6)\\ 
		CMRR  &20\% (30/30) & 27\% (26/28) & 21\%  (22/23) & 19\% (10/10) &  16\% (2/2) &18\% (10/10) & 11\% (6/6)\\
		CMIR & (30/30) & (27/28) &   (23/23) &  (10/10) &   (2/2) & -&-\\
		Open-loop gain  & 16\% (19/30) & 25\% (21/28) & 12\%  (22/23) & 15\% (9/10) &  12\% (2/2) &25\% (9/10) & 11\% (6/6)\\
		Unity-gain bandwidth  &29\% (21/30) & 25\% (22/28) & 19\%  (19/23) & 31\% (5/10) &  19\% (2/2) &32\% (9/10) & 32\% (6/6)\\
		Slew rate &15\% (25/30) & 24\% (15/28) & 16\%  (17/23) & 20\% (10/10) &  26\% (2/2) &22\% (6/10) & 36\% (3/6)\\
		Output voltage swing  & (24/30) & (23/28) &   (19/23) &  (10/10) &   (2/2) & (7/10) &  (4/6)\\\hline
		All {\em Specs} & (6/30) & (8/28) &   (6/23) &  (5/10) &   (0/2) & (2/10)& (2/6)\\\hline
	\end{tabular}
\end{table}

A selection of the outputted  topologies fulfilling all specifications were simulated to analyze how the circuits sized with analytical equations  agree with simulation results (Table~\ref{tab:SizingResults}). A BSIM3v3 transistor models was used for simulations.
Table~\ref{tab:SizingResults} shows the average deviations between analytically calculated and simulated values. The deviations agree well with the expectations of a designer, who expects a deviation below 30\%. 
The largest deviation occurs for the unity-gain bandwidth. Its value is overestimated in the sizing process as it depends linearly on the input conductance of the first stage in the  analytical equations \cite{LakerSansen,ABNG20c}. 

The absolute number of circuits  fulfilling a specification respective all {specs} is also given. For many specification sets, already a number of topologies fulfill all specifications. Other circuits need subsequent optimization \cite{Wicked} to fulfill all specifications.

\section{Outlook: Synthesis of Multi-Stage Op-Amps}\label{sec:OutlookThreeStageOpAmp}
The hierarchical structure of the approach can be extended to other op-amp types, e.g., multistage op-amps.
Fig. \ref{fig:threeStageOpAmp} shows such an op-amp. To synthesize the op-amp and similar topologies automatically following input is derived for Alg.~\ref{algo:SynthesisFunctionalBlock}:
\begin{lstlisting}[mathescape=true]
${S}_1: OP_{so,2, \setminus b_O \setminus cap},$ ${S}_2: A_{inv},$ ${S}_3: cap,$ ${S}_4: cap;$ ${R}_c:$$s_1.out$$\leftrightarrow$$s_2.in_{tc,1},$$s_1.inner_{1, l_{p,st, a_1}} \leftrightarrow s_2.in_{b_{s,1}},$
$s_1.inner_{2, l_{p,st, a_1}} \leftrightarrow$$s_2.in_{b_{s,2}},$$s_1.out_{a_1} \leftrightarrow s_3.plus,$$s_1.out_{a_2}$$\leftrightarrow s_4.plus,$$s_2.out \leftrightarrow$$s_3.minus,$$s_2.out \leftrightarrow s_4.minus;$
$R_f:$$s_1.\Phi_{tc,a_1} \neq s_2.\Phi_{tc},$$s_1.\Phi_{tc,a_2} \neq s_2.\Phi_{tc};$
\end{lstlisting}
The inputted functional blocks are a set of single output two-stage op-amps without their bias circuit $b_O$ and compensation capacitor $cap$, $OP_{so,2, \setminus b_O \setminus cap} = \{A_1, A_2\}$, a set of inverting stages $A_{inv}$, which form the third stage of the op-amps and two sets of capacitors $cap$.
The inverting stage is connected as third stage to the output of the two-stage op-amp.  The stage bias of the third stage is biased by one of the load parts of the first stage $l_{p, st, a_1}$. The capacitors are connected between the outputs of first and third stage, and the outputs of second and third stage. The topologies are only valid if the transconductance   of the third stage has different doping as the transconductance of first and second stage.

\begin{table}[tbp]\centering
	\caption{Results of the sizing method in \protect \cite{ABNG20c} for the three-stage op-amp {\protect Fig. \ref{fig:threeStageOpAmp}} }\label{tab:outputSizingTool}
	\subfloat[Performance values; M: sizing method; S: simulation]{
		\label{tab:performanceValues}
		\setlength{\tabcolsep}{0.1cm}
		\begin{tabu}{|l||l|c|c|}
			\hline
			{Constraints} &{ Spec.} & M & S   \\\hline \hline
			Bias current ($\mu$A) & 10 & 10 & 10\\
			Load Capacity (pF) & 20 & 20 & 20  \\
			Supply voltage (V) & 5 & 5  & 5\\
			Gate-area (10$^3$ $\mu$m$^2$) & $\leq$15  & 8.3& - \\\hline
			Quiescent power (mW) & $\leq$20 & 15 &  17 \\	
			Phase Margin ($^\circ$) & $\geq$60 & - & 50   \\
			CMRR (dB) & $\geq$70& - & 180 \\
			CMIR & 2-3 &  0-4.1  & 0.4-4\\
			Open-loop gain (dB) & $\geq$60 & 88& 73 \\
			Unity-gain bandwidth (MHz) &$\geq$2.5  & 2.8 &1.9 \\\hline
			Slew rate ($\frac{\text{V}}{\mu \text{s}}$)& $\geq$2.5 & 2.6  & 8.4 \\
			Output voltage swing (V) & 1-4   &0.3-4.3 & 0.2-4.2  \\ \hline
		\end{tabu}
	}%
	\qquad
	\subfloat[Device sizes]{
		\label{tab:DimensionsTelescopicOpAmp}
		\setlength{\tabcolsep}{0.1cm}
		\begin{tabu}{|l||l|}
			\hline
			Variable & Value ($\mu$m/fF)\\\hline \hline
			$W_{P_{1,2}};$$L_{P_{1,2}}$ &	6;3 \\
			$W_{P_3}; L_{P_3}$ &	13;6\\
			$W_{P_{4,5}};L_{P_{4,5}}$&	15;1\\
			$W_{P_6}; L_{P_6}$ &	157;9\\
			$W_{P_7}; L_{P_7}$ &256;1\\
			$W_{P_8}; L_{P_8}	$& 203;6\\
			$W_{P_9}; L_{P_9}$ &	104;6\\
			$W_{N_{1,2}};$$L_{N_{1,2}}$ &	319;3\\
			$W_{N_{3,4}};$$L_{N_{3,4}}$&	222;3\\
			$W_{N_5}; L_{N_5}$ &	189;3\\
			$W_{N_6}; L_{N_6}$&	570;1\\
			$W_{N_7}; L_{N_7}$&	37;3\\
			$W_{N_8}; L_{N_8}$&	41;3\\
			$C_1$ &	5\\
			$C_2$ &	100\\
			\hline
		\end{tabu}
	}%
	\qquad
\end{table}

To evaluate the topologies, the sizing method in \cite{ABNG20c} must be adopted to three-stage op-amps.
One of the biggest restrictions in automatic sizing of multi-stage op-amps is the phase margin. The generation of good models is still a research issue in analog design \cite{Shi2}.
The sizing results  for the three-stage op-amp (Table \ref{tab:outputSizingTool}) 
were generated with an enhanced version of  \cite{ABNG20c} without an accurate model for the phase margin.
Instead common stability constraints as in \cite{MultiStage1,MultiStage2} were used. 
The compensation capacitors were manually adopted to improve the stability of the op-amp. 
As the sizing tool provides good values for all specifications not affected by the changed capacitor values  (Table~\ref{tab:outputSizingTool}), the synthesis tool can be well used for the synthesis of multi-stage op-amps. The results will improve when adequate models for the phase margin of multi-stage op-amps  found.

The circuit shown in Fig. \ref{fig:threeStageOpAmp} is a simple version of a three-stage op-amp. Many other types of multi-stage op-amps with more advanced frequency compensation techniques exist \cite{MultiStage1,MultiStage2,Capacitor1, Capacitor2}. They can be  added analogously to the synthesis algorithm increasing the number of supported topologies and functional blocks. 
However,  as the research in this area is still ongoing and an adequate model for the phase margin is missing, this remains future work.

The mentioned multi-stage extension increases the underlying topology library from 3912 to 6852 practically meaningful op-amps. It shows that the library can be easily extended, which is also useful for machine learning-based approaches needing large data sets.

\section{Conclusion}\label{sec:Conclusion}
This paper presented a method for structural synthesis of op-amps by a hierarchical composition graph of functional blocks. The method emulates the manual design process. It uses knowledge described in standard analog design works \cite{Allen,LakerSansen, AnalogIntegratedCircuitDesign, GrayMeyer, Palumbo, MOSCapacitances} and makes  it computationally accessible.  A generic algorithm to compose a functional block from blocks from the same or lower level of abstraction was presented. Rule sets for the hierarchical composition of all functional blocks span a search space of several thousand variants. It is searched by a heuristic mixture of static and dynamic structure exploration and a fast sizing method using behavioral equations \cite{ABNG20c}. 
The method currently supports one- and two-stage op-amps. An outlook is given how also multi-stage op-amps can be supported.

As the functional block representation of op-amps can be similarly encoded as in \cite{Encoding}, future research based on this method can be carried out in the area of machine learning.
Future applications lies in combining the structural synthesis with layout synthesis. 

\begin{acks}
The authors would like to thank the Cusanuswerk for partly funding this work.
\end{acks}

\appendix
\section{Appendix: Implementation Rules for Op-Amp Functional Blocks} 
\label{sec:Rules}

In the following, functional blocks and their composition rules are presented needed to support also symmetrical single-output op-amps (Sec. \ref{sec:FBsymmetricalOpAmps}), fully-differential op-amps (Sec. \ref{sec:FBFullyDiffierential}) and complementary  op-amps (Sec. \ref{sec:FBComplementary}).

\subsection{Symmetrical Op-Amp ($op_{SO,sym}$)} \label{sec:FBsymmetricalOpAmps}
\begin{figure} [tp] \centering
	\includegraphics[width=  0.6 \linewidth]{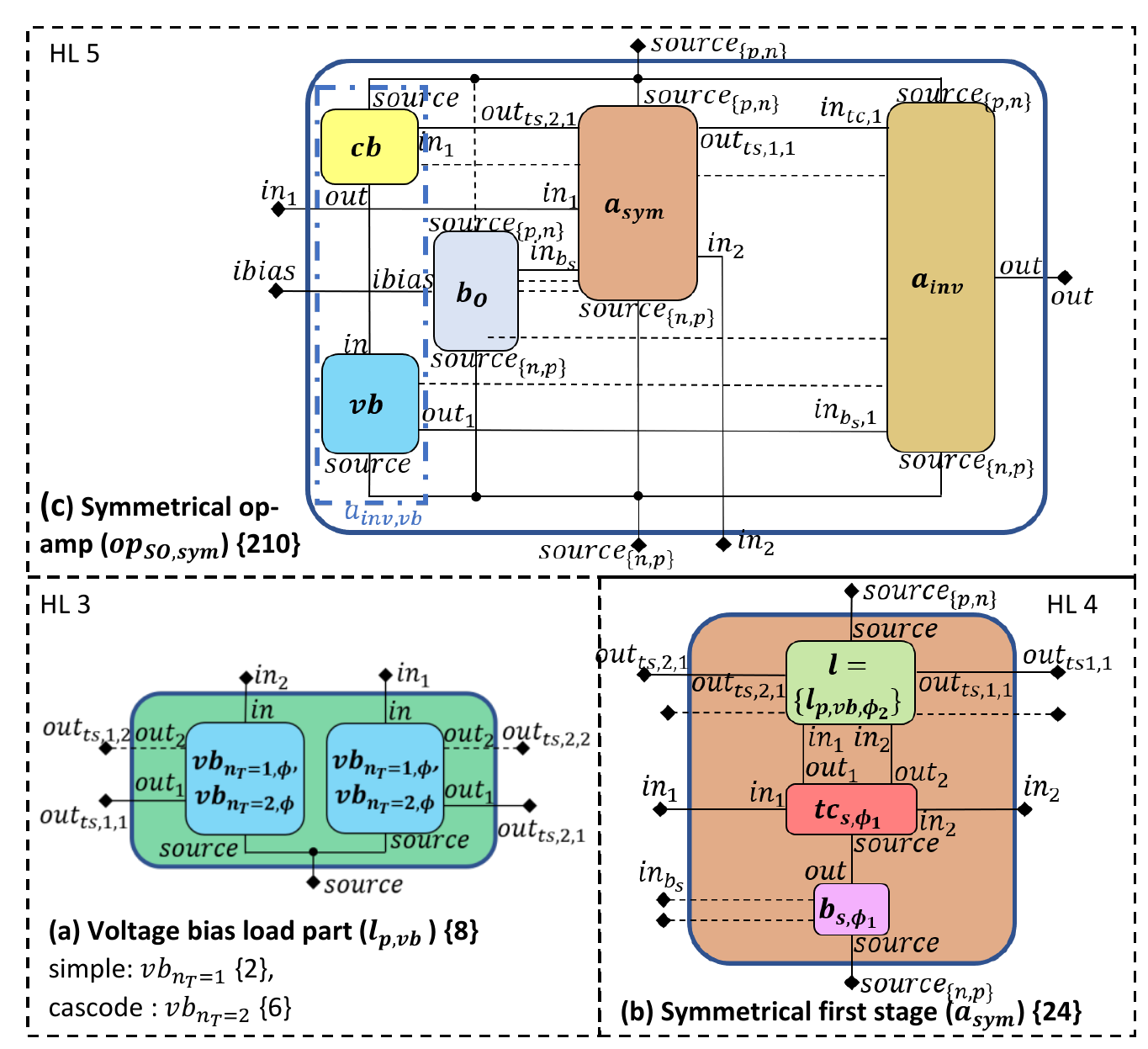}
	\caption{Composition rules for additional functional blocks needed to synthesize symmetrical op-amps, $\{n\}$ denotes the respective number of synthesizable structural implementations}
	\label{fig:FBsSymmetricalOpAmp}
\end{figure}

Fig. \ref{fig:FBsSymmetricalOpAmp} shows the functional blocks and their composition rules for symmetrical op-amps.

{\em Voltage bias load part } implementations ($l_{p, vb}$, Fig.~\ref{fig:FBsSymmetricalOpAmp}a, HL 3) are synthesized based on two identical voltage bias implementations ($vb$, Fig. \ref{fig:symmetricalOpAmp} $P_1 - P_4$) (${S}_1:  VB_{\Phi};$  ${S}_2:  VB_{\Phi}$). The two voltage biases are connected at their sources (${R}_c: s_1 = s_2, s_1.source \leftrightarrow s_2.source$). ${\mathcal{R}}_f: s_1.out_1 \leftrightarrow {s_1}.in$ omits mixed pairs (Fig. \ref{fig:voltageBiasStrucImp}) as voltage bias.

{\em Symmetrical non-inverting stages} ($a_{sym}$, Fig.~\ref{fig:FBsSymmetricalOpAmp}b, HL 4)  feature one load part consisting of voltage biases $l_{p,vb, \Phi_2}$ as load. The outputs of the simple transconductance $tc_s$ are connected to the input pins of the load. 
\begin{lstlisting}[mathescape=true]
${S}_1:  TC_{s, \Phi_1};$  ${S}_2: B_{s, \Phi_1};$ ${S}_3: L = \{L_{p,vb, \Phi_2} \};$ ${R}_c:$$s_1.source \leftrightarrow s_2.out,$$s_1.out_1 \leftrightarrow s_3.in_1,$$s_1.out_2 \leftrightarrow s_3.in_2;$ 
\end{lstlisting}

{\em Symmetrical op-amps ($op_{SO, sym}$)} (Fig.~{\ref{fig:FBsSymmetricalOpAmp}}c, HL 5) are synthesized by adding an inverting stage $a_{inv}$ and an inverting stage with a voltage bias as stage bias $a_{inv,vb}$ to the outputs of a symmetrical non-inverting stage $a_{sym}$. The transconductances $tc_{inv}$, $tc_{inv, vb}$ of the inverting stages $a_{inv}, a_{inv,vb}$ must have the same doping as the load part of the non-inverting stages. The number of transistors must be equal in  $tc_{inv}, tc_{inv,vb}$. The transistor sum in $tc_{inv}, tc_{inv,vb}$  must be greater than or equal to the number of transistors in the load part $l_{p,vb}$ of $a_{sym}$.   The two stage biases $b_{s,inv}, b_{s,inv,vb}$ of $a_{inv}, a_{inv,vb}$  must form a current mirror $cm$.
\begin{lstlisting}[mathescape=true]
${S}_1:  A_{sym, l_{p, vb}.\Phi = \Phi_1};$  ${S}_2: A_{inv, tc_{inv}.\Phi = \Phi_1};$ ${S}_3:$$ A_{inv, vb, tc_{inv}.\Phi = \Phi_1};$${R}_c: n_{T, tc_{inv}} = n_{T, tc_{inv,vb}},$
$n_{T, l_{p, vb}}$$\leq (n_{T, tc_{inv}} + n_{T, tc_{inv, vb}}),$$s_1.out_{ts,1,1} \leftrightarrow s_2.in_{tc,1},$$s_1.out_{ts,2,1} $$\leftrightarrow s_3.in_{tc,1},$$s_2.in_{b_s,1} \leftrightarrow s_3.out{b_s,1},$
$\forall_{n_{T, l_{p, vb}} = 2,n_{T, tc_{inv}} = 2}$$s_2.in_{tc,2} \leftrightarrow s_3.in_{tc,2},$$\forall_{n_{T, l_{p, vb}} = 4, n_{T, tc_{inv}} = 2} [(s_1.out_{ts,1,2}$$\leftrightarrow s_2.in_{tc,2})$$\wedge (s_1.out_{ts,2,2} \leftrightarrow s_3.in_{tc,2}))],$
$n_{T, b_{s,inv}}$$\geq n_{T, b_{s,inv,vb}},$$\forall_{n_{T, b_{s,inv,vb}} = 2}~ s_2.in_{b_s,2} \leftrightarrow s_3.out_{b_s,2};$ ${\mathcal{R}}_f:$$ \mathcal{R}_{f,b_{s,inv},b_{s,inv,vb}} = \mathcal{R}_{f, cm}$
\end{lstlisting}
The bias $b_O$ is generated with Alg. \ref{algo:SynthesisBias}.

\subsection{Fully-Differential Op-Amp ($op_{FD}$)}\label{sec:FBFullyDiffierential}
\begin{figure} [tp] \centering
	\includegraphics[width=  0.6\linewidth]{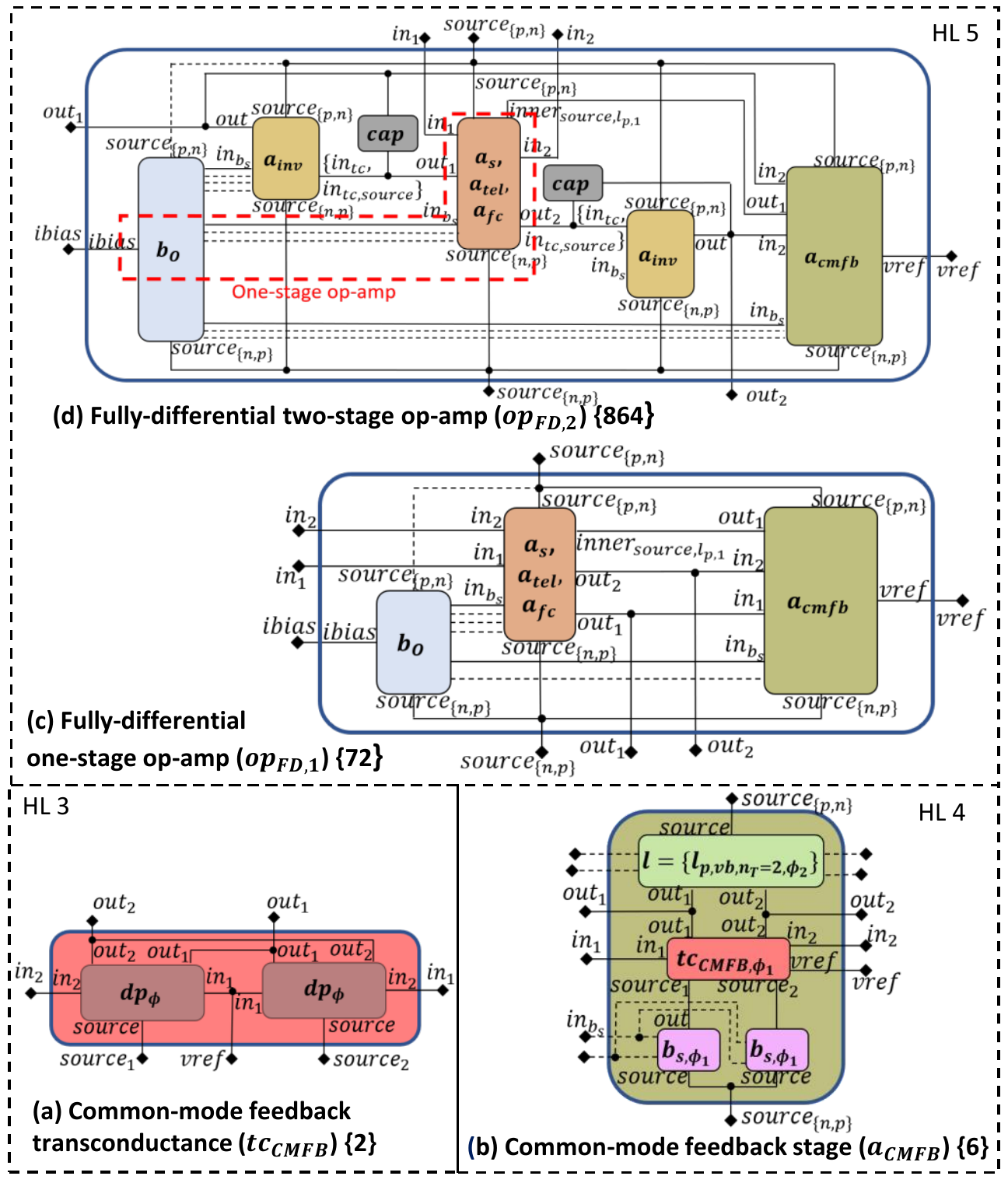}
	\caption{Composition rules for additional  functional blocks needed to synthesize fully-differential op-amps, $\{n\}$ denotes the respective number of synthesizable structural implementations}
	\label{fig:FBsFullyDifferentialOpAmp}
\end{figure}
Fig. \ref{fig:FBsFullyDifferentialOpAmp} shows additional functional blocks and their composition rules for fully differential  op-amps.

{\em Common-mode feedback transconductance} implementations ($tc_{CMFB}$, Fig.~\ref{fig:FBsFullyDifferentialOpAmp}a, HL 3) are generated based on two differential pairs  with equal doping $\Phi$ (${S}_1:  DP_{\Phi};$  ${S}_2: DP_{\Phi}$, e.g, Fig.~\ref{fig:foldedCascodeOpAmp}, $P_8 - P_{11}$). One input of both differential pairs is connected. Also, the outputs are connected (${R}_c: s_1.in_2 \leftrightarrow s_2.in_1,$
$ s_1.out_1 \leftrightarrow s_2.out_2, s_1.out_2 \leftrightarrow s_2.out_1$).

{\em Common-mode feedback stages} ($a_{CMFB}$, Fig. {\ref{fig:FBsFullyDifferentialOpAmp}}b, HL 4) are created based on a common-mode feedback transconductance $tc_{CMFB}$. The load consists of one load part with two simple voltage biases $l_{p,vb, n_T =2}$. Two identical stage biases are connected with their outputs to the respective source pins of the transconductance.   
\begin{lstlisting}[mathescape=true]
${S}_1:  TC_{CMFB, \Phi_1};$ ${S}_2:$ $L =$$\{L_{p,vb, n_T =2, \Phi_2}\};$ ${S}_3: B_{s, \Phi_1};$ ${S}_4: B_{s, \Phi_1};$ 
${R}_c:$ $s_1.out_{1} \leftrightarrow s_2.in_1,$$s_1.out_{2} \leftrightarrow s_2.in_2,$$s_1.source_1 \leftrightarrow s_3.out,$$s_1.source_{2}$$\leftrightarrow s_4.out,$$s_3 = s_4,$
\end{lstlisting}

The algorithm supports one and two-stage fully-differential op-amps.
{\em Fully-differential one-stage op-amp} implementations ($op_{FD, 1}$, Fig.~\ref{fig:FBsFullyDifferentialOpAmp}c, HL 5) are generated based on a first stage and  a feedback stage as input to Alg. \ref{algo:SynthesisFunctionalBlock}. The transconductances of the first stage and the feedback stage have the same doping $\Phi_1$. The outputs of the first stage are  connected to the inputs of the common-mode feedback stage. The output of the feedback stage is connected to the gates of the transistors at the source of the load part of the first stage having the same doping $\Phi_2$ as the load of the feedback stage. 
\begin{lstlisting}[mathescape=true]
${S}_1: A_{s, tel, fc, tc_{ninv}.\Phi = \Phi_1};$ ${S}_2: A_{CMFB, tc_{ninv}.\Phi = \Phi_1};$ ${R}_c: s_1.out_1 $$\leftrightarrow s_2.in_1,$$s_1.out_2 \leftrightarrow s_2.in_{2},$$s_1.inner_{1,l_{p, \Phi_2}} \leftrightarrow s_2.out_1;$
\end{lstlisting}
The bias $b_O$ is generated with Alg. \ref{algo:SynthesisBias}.
	
{\em Fully-differential two-stage op-amps} ($op_{FD, 2}$, Fig.~\ref{fig:FBsFullyDifferentialOpAmp}d, HL~5) are synthesized by adding an inverting stage and a capacitor to each output of a first stage. The inverting stages are symmetrical. Their outputs are input to the feedback stage. The output of the feedback stage is fed to the gates of the transistors at the source of the load part of the first stage having the same doping as the load of the feedback circuit. The transconductances of the feedback circuit and the first stage have the same doping. The input to Alg. \ref{algo:SynthesisFunctionalBlock} is:
\begin{lstlisting}[mathescape=true]
${S}_1:   A_{s, tel, fc,  tc_{ninv}.\Phi = \Phi_1};$ ${S}_2: cap;$ ${S}_3: cap;$ ${S}_4: A_{inv};$ ${S}_5: A_{inv};$  ${S}_6: A_{CMFB, tc_{ninv}.\Phi = \Phi_1};$  
${R}_c: s_4 = s_5 ,$$s_1.out_1 $$\leftrightarrow s_2.plus, $$s_1.out_2 \leftrightarrow s_3.plus, $$s_1.out_1 \leftrightarrow s_4.in_{tc,1}, $$s_1.out_2 $$\leftrightarrow s_5.in_{tc,1},$$s_2.minus \leftrightarrow s_4.out, $
$s_3.minus \leftrightarrow s_5.out, $$s_2.out \leftrightarrow s_6.in_1, $$s_3.out \leftrightarrow s_6.in_{2}, $$s_1.inner_{1,l_{p, \Phi_2}} \leftrightarrow s_6.out_1; $
\end{lstlisting}
The bias $b_O$ is generated with Alg. \ref{algo:SynthesisBias}.
	
\subsection{Complementary Op-Amp ($op_{comp}$)}\label{sec:FBComplementary}

\begin{figure} [tp] \centering
	\includegraphics[width=  0.6\linewidth]{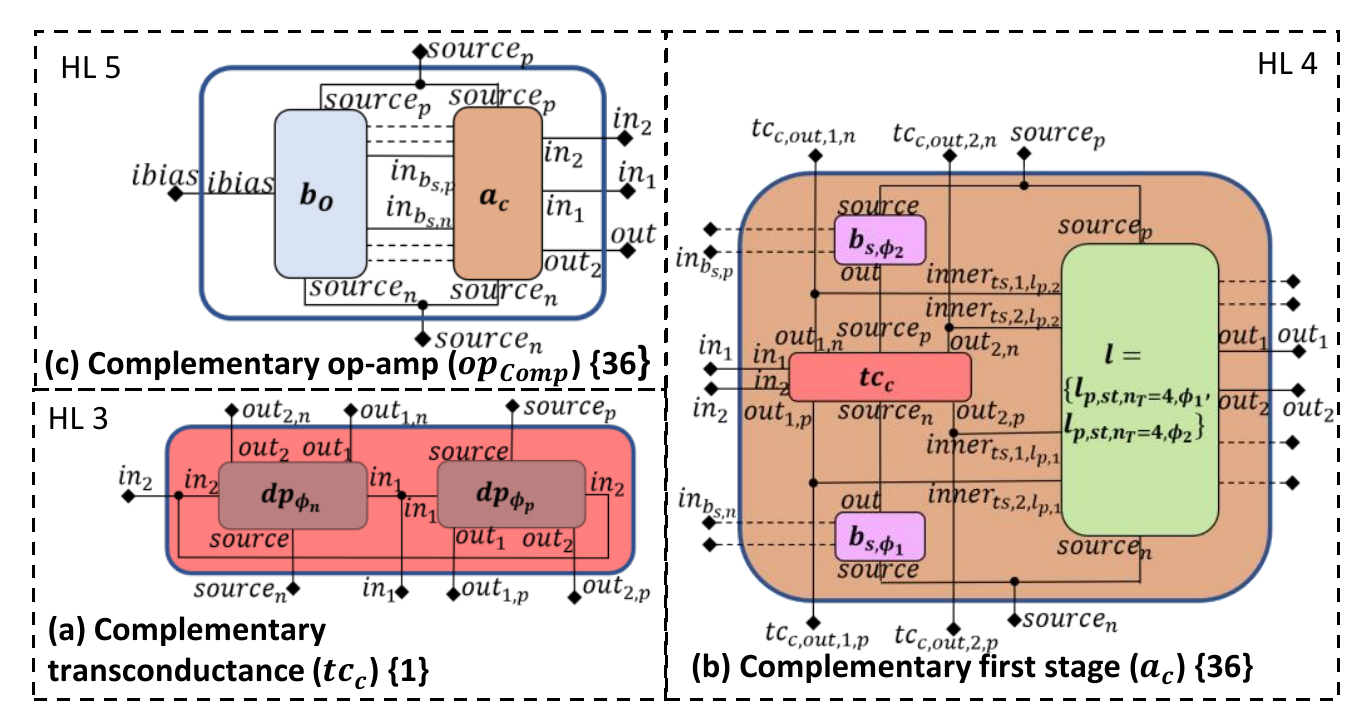}
	\caption{Composition rules for additional functional blocks needed to synthesize complementary op-amps, $\{n\}$ denotes the respective number of synthesizable structural implementations}
	\label{fig:FBsComplementaryOpAmp}
\end{figure}
Fig. \ref{fig:FBsComplementaryOpAmp} shows additional functional blocks and their composition rules for complementary op-amps.

{\em Complementary transconductances} ($tc_c$, Fig. {\ref{fig:FBsComplementaryOpAmp}}a, HL 3) are synthesized based on two differential pairs (${S}_1:  DP_{\Phi_1}, {S}_2: DP_{\Phi_2}$) having different dopings $\Phi_1, \Phi_2$ (Fig. \ref{fig:railToRailAmplifier}, $N_1, N_2, P_1, P_2$). Both inputs of the differential pairs are connected (${R}_c: s_1.in_1 \leftrightarrow s_2.in_1,$
$s_1.in_2 \leftrightarrow s_2.in_2$).

{\em Complementary first stages} ($a_{c}$, Fig.~{\ref{fig:FBsComplementaryOpAmp}}b, HL 4) are created based on  a complementary transconductance $tc_c$. As $tc_c$ has two source pins of different doping, two stage biases $b_{s, \Phi_1}, b_{s, \Phi_2}$ of different doping are connected with their outputs to the sources of the same doping. The stage biases should be symmetrical, i.e., have the same structural implementation by different doping.
The load is a two-load-part load with eight transistors. The inner pins $inner_{ts_i, l_{p,j}}$ of current or voltage biases in the load are connected to the output of the transconductance having a different doping as the load part.
\begin{lstlisting}[mathescape=true]
${S}_1:  TC_{c};$  ${S}_2: B_{s, \Phi_1};$ ${S}_3: B_{s, \Phi_2};$ ${S}_4: L_{n_{l_p} = 2}$ ${R}_c:$$\text{sym}(s_2, s_3),$$s_1.source_{\Phi_1} \leftrightarrow s_2.out,$$s_1.source_{\Phi_2} \leftrightarrow s_3.out,$
$s_1.out_{1, \Phi_1} \leftrightarrow s_4.inner_{ts_1, l_{p,2}},$$s_1.out_{2, \Phi_1} \leftrightarrow s_4.inner_{ts_2, l_{p,2}},$$s_1.out_{1, \Phi_2} \leftrightarrow s_4.inner_{ts_1, l_{p,1}},$$s_1.out_{2, \Phi_2} \leftrightarrow s_4.inner_{ts_2, l_{p,1}};$
\end{lstlisting}

{\em Complementary op-amps ($op_{Comp}$)}  (Fig. {\ref{fig:functionalBlockLibrary}}s) are synthesized using complementary first stages 
(${S}_1$:  $A_{c}$). The topology specific op-amp bias is created with Alg. \ref{algo:SynthesisBias}.

\bibliographystyle{ACM-Reference-Format}
\bibliography{custom_added_2}

\end{document}